\documentclass[aps,prb,showpacs,preprintnumbers,twocolumn,superscriptaddress]{revtex4-2}
\usepackage{amsmath,amssymb}
\usepackage{bm}
\usepackage{tipa}
\usepackage{upgreek}
\usepackage{comment}
\usepackage{mathrsfs}
\usepackage{graphicx}
\usepackage{braket}
\usepackage{enumitem}
\usepackage{mathbbol}
\usepackage{booktabs}
\usepackage{gensymb}
\usepackage[normalem]{ulem}
\usepackage{color}
\usepackage[colorlinks,bookmarks=true,citecolor=blue,linkcolor=red,urlcolor=blue]{hyperref}
\usepackage{hyperref}

\usepackage{pifont}

\newcommand{\bk}{\bm{k}}

\renewcommand{\comment}[1]{}
\usepackage{subfigure}

\begin{document}

\title{Tuning between a fractional topological insulator and competing phases at $\nu_\mathrm{T}=2/3$}
\author{Roger Brunner}
\affiliation{Institute for Theoretical Physics, ETH Z{\"u}rich, 8093 Z{\"u}rich, Switzerland}
\author{Titus Neupert}
\affiliation{Department of Physics, University of Z{\"u}rich, Winterthurerstrasse 190, 8057 Z{\"u}rich, Switzerland}
\author{Glenn Wagner}
\affiliation{Institute for Theoretical Physics, ETH Z{\"u}rich, 8093 Z{\"u}rich, Switzerland}

\begin{abstract}
We study a spinful, time-reversal symmetric lowest Landau level model for a flatband quantum spin Hall system at total filling fraction $\nu_\mathrm{T}=2/3$. Such models are relevant, e.g.,~for spin-valley locked moiré transition metal dichalcogenides. The opposite Chern number of the two spins hinders the formation of a quantum Hall ferromagnet, instead favouring other phases. We study the phase diagram in dependence on different short-range Haldane pseudopotentials $v_m$ and uncover several phases: A fractional topological insulator, a phase separated state, a spin-polarized fractional quantum Hall state, and the partially particle-hole transformed Halperin (111) state. The effect of the pseudopotentials $v_m$ depends on the parity of $m$, the relative angular momentum. 
\end{abstract}

\maketitle

\section{Introduction}

Moiré materials are distinguished by their remarkable tunability. Due to their two-dimensional nature, electrostatic doping can be used to tune the density. Moreover, the constituent layers can be chosen from a broad variety of stackable two-dimensional van-der-Waals materials, offering access to a wide range of underlying band structures \cite{Geim2013}. Crucially, these band structures can be further engineered through control over twist angle, displacement field, substrate choice (which may break inversion symmetry, as in the case of hexagonal boron nitride, or induce spin-orbit coupling, as in transition metal dichalcogenides), and perturbations such as strain. Strain, in particular, has a pronounced influence on the electronic structure of moiré systems \cite{Parker2021,Kwan2021,Wagner2021b,Wang2021}. While in-situ control over strain remains limited, spatial variations naturally present in fabricated devices can be imaged with high resolution \cite{Choi2019,Kerelsky:2019aa,Xie:2019aa}, allowing for the identification of regions with desirable strain profiles \cite{Nuckolls2023}. These capabilities illustrate the significant versatility available for tailoring the single-particle band structure in moiré heterostructures. However, the many-body physics of these systems depends equally on the interaction term in the Hamiltonian, whose controllability is of growing experimental and theoretical interest.

Engineering electron-electron interactions is a key step toward realizing novel correlated and topological phases. One of the most direct means of tuning interactions is through screening. In twisted bilayer graphene, for example, the distance to metallic gates has been employed as a handle to modulate screening \cite{Stepanov2020,Saito2020}. Proximity to a tunable metallic layer presents another avenue \cite{Liu2021}, as does the use of substrates with high and controllable dielectric constants. Strontium titanate, with its tunable dielectric constant reaching values on the order of 10000, exemplifies this approach and offers in-situ tunability \cite{gao2024doubleedgedroleinteractionssuperconducting}. Additionally, multilayers of exfoliable high-dielectric materials such as HfO$_2$ can be used to engineer the spatial profile of the interaction potential \cite{kwan2024abelianfractionaltopologicalinsulators}. Interactions are also influenced by electron-phonon coupling, which may mediate effective attraction and influences the phase diagram of twisted bilayer graphene \cite{kwan2023electronphonon,Liu2024}. This coupling is sensitive to substrate alignment and thus experimentally tunable \cite{Chen2024}.

In future experiments, interaction engineering promises to elevate moiré materials into a regime akin to tunable quantum simulators, as realized in cold atom systems~\cite{Gross2017}. Among the most intriguing theoretical targets is the fractional topological insulator (FTI)—a time-reversal (TR) symmetric state exhibiting both topological order and fractionalization \cite{bernevigQSHE2006,Levin_Stern,Neupert2011FTI,Stern2015review,Neupert_2015,Levin2012classification}. This state realizes the long-sought fractional quantum spin Hall effect. Moiré systems offer a promising platform for realizing this elusive phase, but interaction control is essential for its stabilization \cite{kwan2024abelianfractionaltopologicalinsulators}.

\begin{figure}[b]
    \centering
    \includegraphics[width=\linewidth]{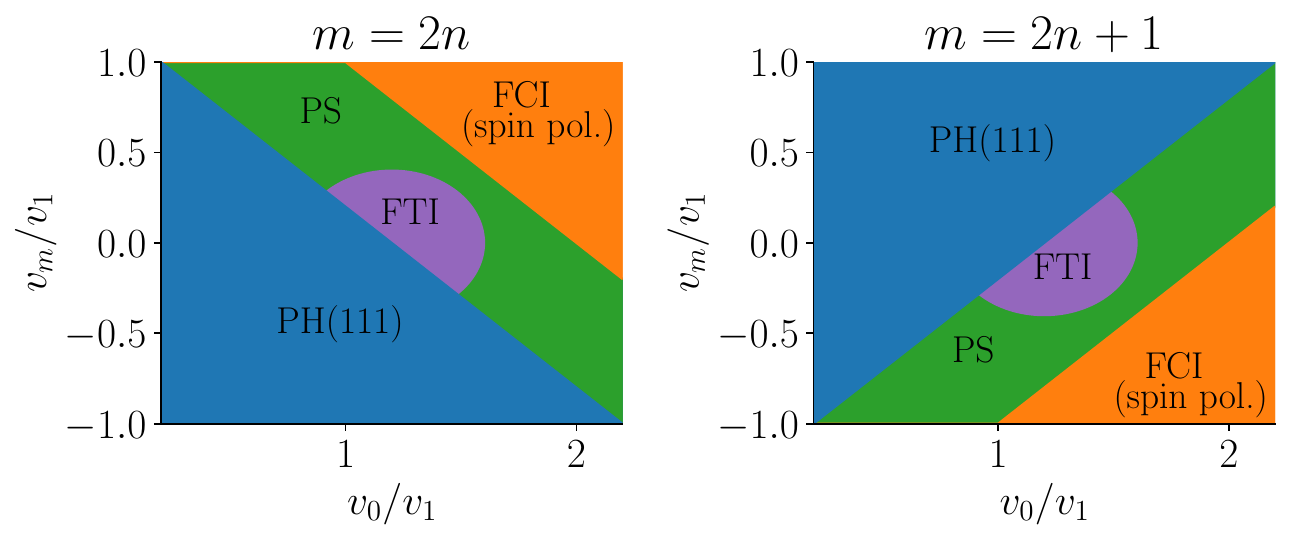}
    \caption{Schematic phase diagram at $\nu_\textrm{T}=2/3$ as a function of Haldane pseudopotentials $v_0/v_1$ and $v_m/v_1$ for even (left) and odd (right) angular momentum $m$. The phases listed in the diagram are a partially particle-hole transformed Halperin 111 state PH(111), a fractional topological insulator (FTI), phase separation (PS) and a spin polarized fractional Chern insulator (FCI). }
    \label{fig:intro_sketch}
\end{figure}

Twisted transition metal dichalcogenide heterostructures, and specifically twisted MoTe$_2$, host pairs of TR related bands near charge neutrality for small twist angles \cite{Wu2019}. These bands are flat, of opposite Chern number and contain electrons of opposite spin flavour due to the spin-orbit coupling induced spin-valley locking. This band structure at small twist angles $\theta\sim2^\circ$ has been proposed to be very close to a Landau level description \cite{Ahn_2024}, allowing for a simplified modeling of the system using TR symmetric Landau level ``bilayers''. The layers correspond to the spin index that by TR symmetry directly matches the sign of the Chern number associated with each spin species through $\mathrm{sgn}(\sigma_z)=\mathrm{sgn}(C)$ \cite{Neupert_2015,Ahn_2024}. (Note that unlike physical bilayers the spin species are not spatially separated.) Such quantum Hall systems with additional degrees of freedom are known to exhibit exotic phases \cite{Moon1995,Ezawa_2009, Sondhi1993,Barkeshli2010}. Unlike the model of interest above, however, where the two spin species correspond to opposite Chern numbers $C$, the multicomponent systems \cite{girvin1995multicomponentquantumhallsystems,QHB_review} mostly studied so far consist of bands of equal Chern number. The magnetic field setup hinders TR symmetric phases and, as a consequence, Landau levels of electrons of both spin species with $(C_\uparrow,C_\downarrow)=(1,-1)$ were not investigated beyond a few initial studies \cite{Chen2012FQHtorusFTI,furukawa2014global,Repellin2014FTI}. However, with the newly gained access to TR related Chern bands in transition metal dichalcogenide band structures, there have been many developments toward understanding the opposite Chern setup \cite{Mukherjee2019FQHsphereFTI,Bultinck2020mechanism,zhang2018composite,kwan2021exciton,kwan2022hierarchy,eugenio2020DMRG,stefanidis2020excitonic,chatterjee2022dmrg,myersonjain2023conjugate,yang2023phase,Wu2024,shi2024excitonic,kwan2024textured,abouelkomsan2023band,kwan2024abelianfractionaltopologicalinsulators,kwan2024abelianfractionaltopologicalinsulators, wagner2025variationalwavefunctions}.
Within this configuration, fractional Chern insulators (FCIs) \cite{neupert,sheng,regnault} and fractional topological insulators might arise, where the former have already been realized experimentally in twisted MoTe$_2$ \cite{cai2023signatures,zeng2023integer,Park2023,Xu2023,park2024ferromagnetism,xu2024interplay}. As of yet, fractional topological insulators have been experimentally elusive. 

An exact diagonalization study of a TR invariant lowest Landau level (LLL) revealed that the existence of the FTI depends strongly on interaction amplitudes, specifically the suppression of the onsite part of the Coulomb interactions. Further, they showed that the long-range interactions for $\nu_\uparrow=\nu_\downarrow=1/3$ filling are beneficial to the FTI \cite{kwan2024abelianfractionaltopologicalinsulators}. This indicates a clear departure from the limit of two decoupled FCIs, whose ground state is the tensor product of TR conjugated zero-energy eigenstates of the Laughlin parent Hamiltonian, i.e.~the Laughlin state $\ket{\frac{1}{3}}\otimes {\ket{\bar{\frac{1}{3}}}}$ \cite{laughlin1983anomalous}. In Ref.~\onlinecite{wagner2025variationalwavefunctions} it was found that there are no zero energy eigenstates of the coupled $\nu_\mathrm{T}=2/3$ TR symmetric quantum Hall Hamiltonian and the only tractable parent Hamiltonian is the above limit of two decoupled Laughlin Hamiltonians, despite complete absence of inter-spin correlations. This Hamiltonian only exhibits short-range intra-spin correlations, but has been shown to be stable for small short-range inter-spin interactions, before transitioning into a phase separated state upon increased amplitude \cite{Chen2012FQHtorusFTI}. Phase separation is also observed in Refs.~\onlinecite{Mukherjee2019FQHsphereFTI,crepel2024attractive}, where the latter reference additionally studied the competition between the FTI and a superconducting phase. 

\begin{figure*}[htbp!]
    \centering
    \subfigure[]{\includegraphics[width = 2.3in]{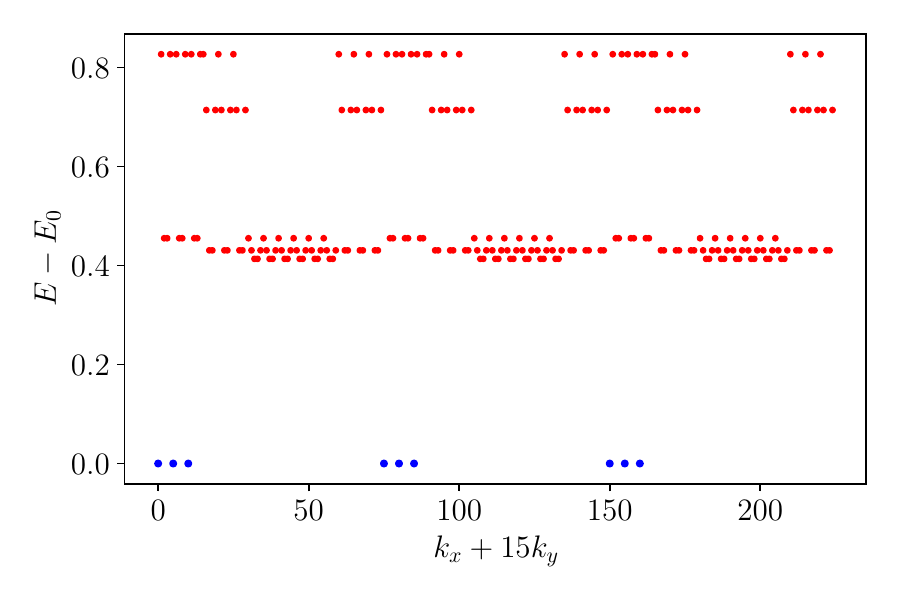}}
    \subfigure[]{\includegraphics[width = 2.3in]{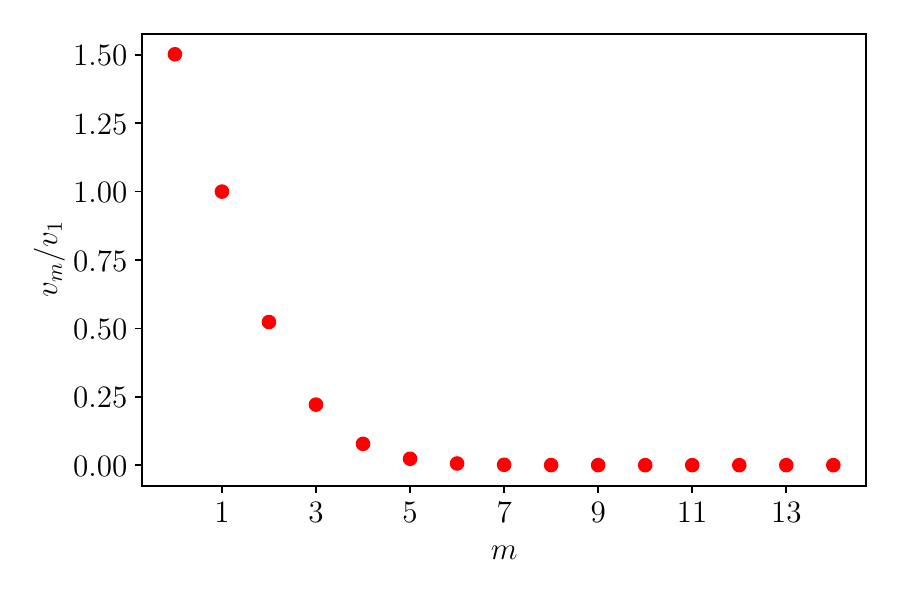}}
    \subfigure[]{\includegraphics[width = 2.3in]{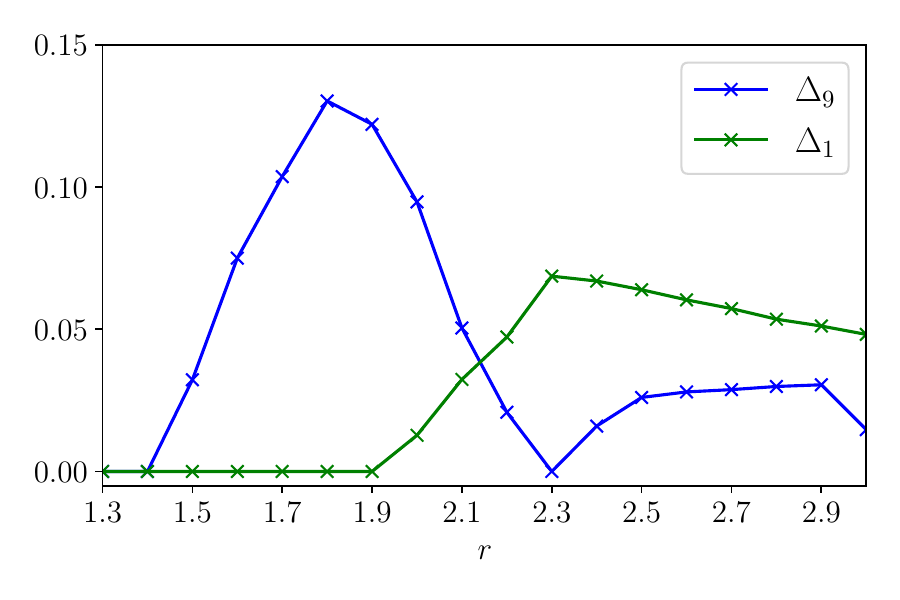}}
    
    \caption{(a) Spectrum of the reference state of two decoupled Laughlin states on the torus for the system size $N_\Phi=15$ at $\nu_\uparrow=\nu_\downarrow=1/3$, where $N_\Phi$ is the number of flux quanta that thread the torus. The nine degenerate ground states are marked in blue colour and aid in identifying the ground state momentum sectors of the FTI. (b) Numerical values for the normalized pseudopotentials $v_m/v_1$ as a function of relative angular momentum for a hard-core interaction $V_{\mathrm{HC}}=\Theta(R-|z|)$, where $\Theta(z)$ is the Heaviside step function, $z$ are the relative holomorphic coordinates of the two particles and $R$ is the range of the interaction. The radius is measured by the dimensionless quantity $r=R^2/4l_B^2$, where $l_B$ is the magnetic length. We show the pseudopotentials for $r=1.9$. (c) The gap size $\Delta_\mathrm{GSD}=E_\mathrm{GSD}-E_\mathrm{GSD-1}$ for the FTI at $\mathrm{GSD}=9$ (blue) and the PH(111) state at $\mathrm{GSD}=1$ (green) as a function of the distance parameter $r$ for the fermionic hard-core interaction defined in (b). The phase is identified according to the $\mathrm{GSD}$ featuring the largest gap.}
    \label{fig:1}
\end{figure*}

These findings motivate the investigation of phase diagrams in terms of pseudopotential interactions to find ways to stabilize FTIs  and illuminate the competing phases. To achieve this, we draw on the description of the LLL in the symmetric gauge and investigate the interaction Hamiltonian in terms of its coefficients, the Haldane pseudopotentials $v_m$, where $m$ is the relative angular momentum between the two interacting particles \cite{Haldane1983}. We use the pseudopotential formalism to define toy Hamiltonians at $\nu_\uparrow=\nu_\downarrow=1/3$ filling featuring few active pseudopotentials in order to investigate their individual action on the TR symmetric LLL. They manifest in the nature of the ground state for given pseudopotential configuration which may change upon variation of the amplitudes. This information can be gathered in $2$-dimensional phase diagrams in terms of the on-site and a single long-range pseudopotential $(v_0,v_m)$. We observe the FTI in the phase diagrams for all $m$. Furthermore, an even-odd effect in $m$ is apparent. Phase diagrams with equal $m$-parity are similar to each other, but different for opposite parities, suggesting a qualitative equivalence $v_{2n}\sim -v_{2n+1}$. We show the schematic phase diagram in Fig.~\ref{fig:intro_sketch}.


\section{Methods}


We investigate a spinful $\nu_\mathrm{T}=2/3$ filled lowest Landau level where the two spin species carry opposite Chern number. The kinetic energy is fully quenched by virtue of being in the LLL. The model interaction Hamiltonian is built by defining the Haldane pseudopotentials
\begin{equation}
    v_m = \bra{m}\hat{V}\ket{m},
    \label{eq: Haldane PP}
\end{equation}
where $\ket{m}$ are two-body eigenstates of the LLL Hamiltonian in the relative angular momentum $m>0$ basis. The Hamiltonian  can be written using projectors $\mathcal{P}_m^{\sigma\sigma'}$ into states of relative angular momentum $m$ and spins $\sigma$ and $\sigma'$ as
\begin{equation}
    H_{\mathrm{int}}=\sum_{m,\sigma}v_m\mathcal{P}_m^{\sigma\sigma} + \frac{1}{2}\sum_{m,\sigma,\Bar{\sigma}}v_m\mathcal{P}_m^{\sigma\Bar{\sigma}},
    \label{eq: Hamil}
\end{equation}
where $\Bar{\sigma}$ indicates the opposite spin species to $\sigma$. Due to fermionic anti-symmetry, intra-spin interactions only consist of odd $m$ pseudopotentials, whereas inter-spin interactions contain the full set $v_{\{m\}}$.
 
Using exact diagonalization, we investigate the model on the torus with aspect ratio $1$ and an angle $\pi/2$ between the two axes. The system size is determined by $N_\Phi=15$ flux quanta threading the torus, with $N_e=10$ electrons. We restrict to the LLL and neglect Landau level mixing. However, we note that band mixing in twisted MoTe$_2$ cannot always be neglected and is often relevant in realistic systems~\cite{kwan2024abelianfractionaltopologicalinsulators}. The torus manifests the ground state degeneracy ($\mathrm{GSD}$) of topologically ordered states in the $E(\bk)$ spectra, where $\bk=(k_x,k_y)$ is the many body momentum of the system \cite{regnault}. For the system size of $N_\Phi=15$ flux quanta, the momentum components take values in $\{0,\dots,14\}$, due to periodic boundary conditions in both directions. For a fully spin polarized system, additional symmetries reduce the many body momentum phase space as discussed in appendix \ref{App1}.

The parent Hamiltonian for the FTI in the limit of decoupled FCIs has a single intra-spin pseudopotential $v_1>0$. The corresponding spectrum exhibits an exact nine-fold degenerate ground state. This degeneracy can be captured by the FTI quality parameter $s_9/\Delta_9$, which measures the maximal energy difference between neighboring ground states, i.e. the spread $s_9=\max(E_{i+1}-E_i)$, $i\in\{0,\dots,8\}$, and compares it to the spectral gap that separates the ground state manifold from the excited states, $\Delta_9=E_9-E_8$, where $E_i$ are eigenstate energies ordered by magnitude. In the above limit, the degeneracy of the ground states is exact such that $s_9=0$ and, thus, the FTI quality parameter is zero, implying an FTI of optimal quality [Fig. \ref{fig:1} (a)]. Long-range interactions induce departure from the ideal FTI limit and result in nonzero $s_9$ up to total loss of topological order when the gap closes at $s_9/\Delta_9\sim1$.

\begin{figure*}
    \centering
    \includegraphics[width = \linewidth]{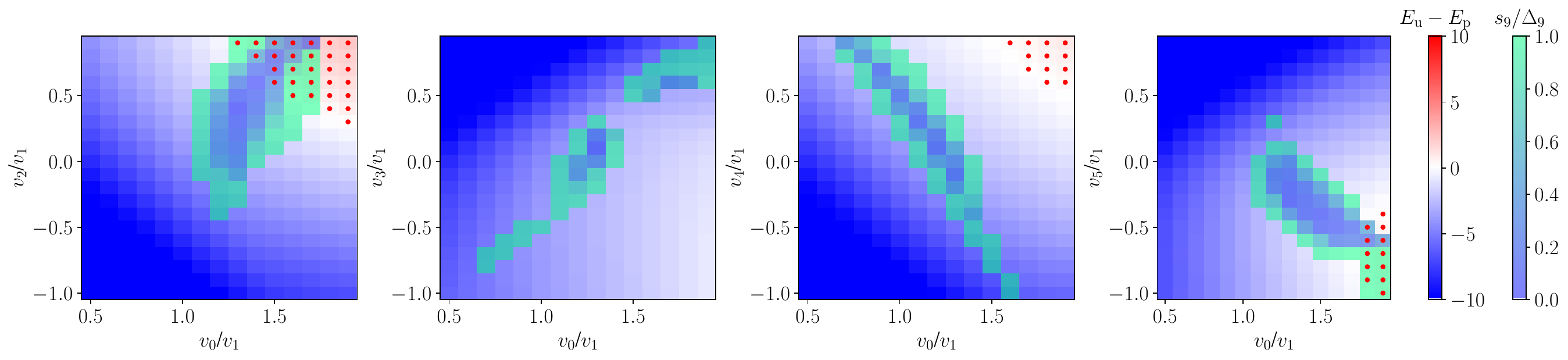}
    \caption{Phase diagrams for the ground state spin polarization as a function of $v_m$ and $v_0$ (for fixed $v_1$) measured by the difference between the ground state energies of the spin unpolarized and spin polarized sectors, $E_\mathrm{u}-E_\mathrm{p}$. The system parameters are $N_\Phi=15$ and $N_e=10$. The red dots mark spin polarized regions. Inside these regions, an FCI with $3$-fold degeneracy arises. The FTI quality parameter $0\leq s_9/\Delta_9 < 1$ (green-blue regions) is shown to illustrate its alignment with equi-energy-difference lines. Competition with full spin polarization appears to be beneficial to the FTI. Consideration of the full set of pseudopotentials $v_m$ in appendix \ref{App0} reveals an apparent even-odd effect with respect to the favoured polarization at repulsive or attractive interactions, which is owed to the fact that $v_{2n}$ are purely inter-spin interactions and hence the energy penalty for repulsive interactions can be avoided by polarization. For $v_{2n+1}$, which include intra-spin interactions, polarization is favoured for attractive interactions.}
    \label{fig:2}
\end{figure*}
\begin{figure*}
    \centering
    \includegraphics[width = \linewidth]{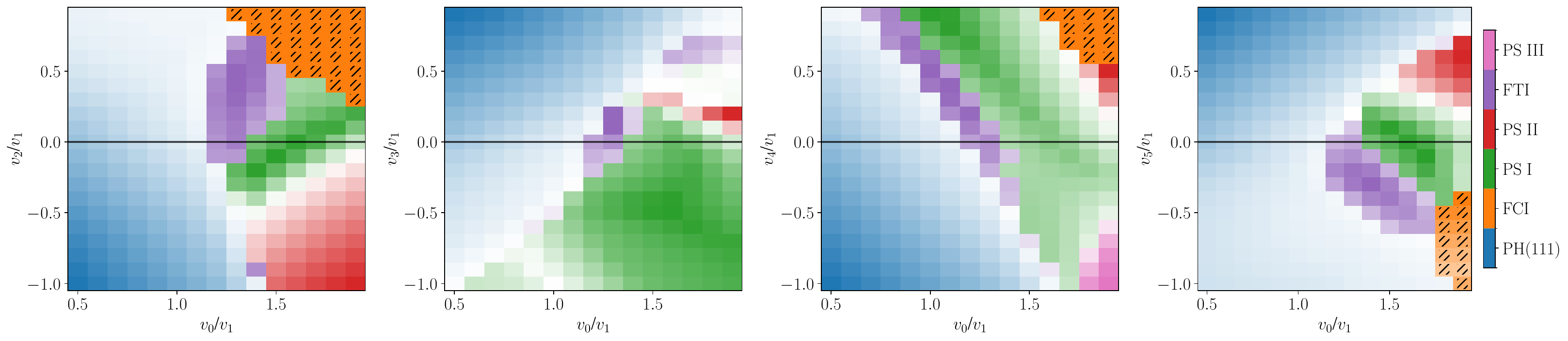}
    \caption{Phase diagrams for the ground state degeneracy in terms of individually tuned $v_m$ and $v_0$ for fixed $v_1$. The system parameters are again $N_\Phi=15$ and $N_e=10$. The colors determine the phases according to their respective $\mathrm{GSD}$ and the transparency reflects the size of the quality parameter $\Delta_{\mathrm{GSD}}/\max_{\{(v_0,v_m)\}}\Delta_\mathrm{GSD}$, being fully opaque when it is one. Hatched regions indicate full spin polarization and white regions reflect spectra for which a $\mathrm{GSD}$ cannot be identified conclusively or are gapless. Featured in the phase diagrams are the partially particle-hole transformed Halperin (111) state ($\mathrm{GSD}=1$), the FCI in the polarized region ($\mathrm{GSD}=3$), the FTI ($\mathrm{GSD}=9$) and phase separated states, i.e. $\mathrm{PS}$ I-III, of different ground state momentum sectors ($\mathrm{GSD}\mod 4 =0$). Details of the phases in terms of composition and symmetry induced degeneracies are given in table \ref{tab:1}. To discriminate against false detection of ground state degeneracies during transitions, the condition of featuring the correct momentum sectors was added. Upon inspection of the full set of pseudopotentials $v_m$ in appendix \ref{App0}, there is an evident even-odd effect in $m$, which dictates the direction and extent of the FTI phase in the phase diagram.}
    \label{fig:3}
\end{figure*}

\begin{figure}[htpb!]
    \centering
    \includegraphics[width=\linewidth]{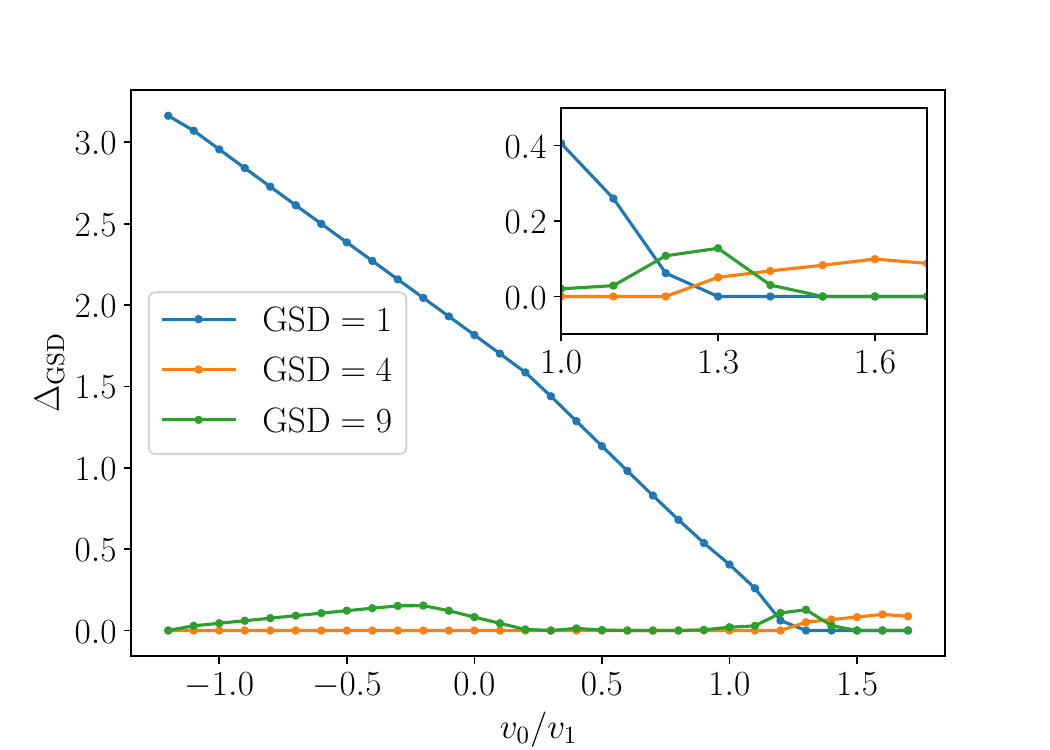}
    \caption{The gap size of given ground state dimension $\Delta_\mathrm{GSD}=E_{\mathrm{GSD}+1}-E_\mathrm{GSD}$ for a system where $v_{m\geq2}=0$ along $v_0/v_1$ into large attractive amplitudes (right to left). The single ground state phase at moderate repulsive interactions $0<v_0/v_1\lesssim 1.2$ is clearly connected to the state at large negative $v_0$. The gaps for $\mathrm{GSD}=4$ and $9$ are added to illustrate the phase transitions for large repulsive on-site interactions. In the inset, the region of $v_0$ relevant to the FTI is shown more closely to illustrate the similar gap size of the FTI and phase separated phase and where the transitions occur.}
    \label{fig:4}
\end{figure}

 Phases without topological order exhibiting spontaneous spatial symmetry breaking feature multiple ground states that are exactly degenerate (provided the order is classical, in that the order parameter commutes with the Hamiltonian). For such phases, the size of the quality parameter defined above is not informative as it vanishes by symmetry. Also for phases with a single ground state the quality parameter has to be adapted as the ground state manifold does not feature a spread. Instead, we therefore use the size of the gap as the identifying property and assign the dimension of the ground state manifold according to the maximal gap $\max_{\mathrm{GSD}} \Delta_{\mathrm{GSD}}=\max_{\mathrm{GSD}}(E_{\mathrm{GSD}+1}-E_{\mathrm{GSD}})$. Fig.~\ref{fig:1} (b) displays the Haldane pseudopotentials for a hard-core interaction $V_{\mathrm{HC}}=\Theta(R-|z|)$, where $\Theta(z)$ is the Heaviside step function and $z$ are the relative holomorphic coordinates of the two particles and $R$ is the range of the interaction. Using the method of largest gap, $\mathrm{GSD}=9$ is identified for a large range of interaction ranges, indicating the FTI [Fig. \ref{fig:1} (c)]. The FTI is accompanied by a phase that exhibits a single ground state upon increase of $R$, which  acts to suppress on $v_0$.

In some cases, especially for phase diagrams containing multiple phases, the quality parameter has to be extended further for a consistent measure of stability inside the phases. After the identification of the ground state degeneracy for given pseudopotential configuration according to $\max_{\mathrm{GSD}} \Delta_{\mathrm{GSD}}$, the quality parameter is then given by the relative gap size within given phase, i.e. $\Delta_{\mathrm{GSD}(\{v_m\})}/\max_{\{v_m\}}\Delta_{\mathrm{GSD}(\{v_m\})}$. The phase is therefore the most stable where its quality parameter takes the value $1$ and the instability is measured by how quickly the quality parameter falls off towards a transition. As this method also captures accidental ground state degeneracies, the spectra are inspected by eye and only phase-specific $\mathrm{GSD}$s are taken into account.


\section{Results}


 Our main findings are presented in Fig.~\ref{fig:2} and Fig.~\ref{fig:3}, which show the ground state spin polarization and ground state degeneracy, respectively. The plots are computed for a system with the three active pseudopotentials $v_0$, $v_1=1$ and $v_m$ with $m\in\{2,3,4,5\}$. The diagrams for the remaining $v_m$ with $m\in\{6,\dots,14\}$ are featured in appendix~\ref{App0}.
 
 In Fig.~\ref{fig:2} we observe that large repulsive $v_{2n}$ pushes the system to spontaneously spin polarize to avoid energy penalties. For attractive $v_{2n}$, the spin unpolarized case is favoured. It is worth noting that the $v_{2n}$ pseudopotentials are purely inter-spin interactions due to fermion statistics. The $v_{2n+1}$ pseudopotentials, however, contribute as both intra-spin and inter-spin interactions and hence the system prefers spin polarization for attractive $v_{2n+1}$ and the unpolarized state for repulsive $v_{2n+1}$. This reflects the fact that for given $v_{2n+1}$ the intra-spin interactions dominate over the inter-spin interactions due to the form factors of electrons of opposite Chern number being smaller. For this reason, there is an energy penalty for spin polarization for large repulsive $v_{2n+1}$. 
Interestingly, even though the FTI requires a vanishing spin polarization, the FTI quality parameter is smallest close to the phase boundary between the spin polarized and spin unpolarized region.

\begin{table}[htpb!]
    \centering
    \begin{tabular}{c|c||c|c|c||c}
         Phase  & Spin Pol. & $\lambda_{C_4}$ & $\lambda_{M_y}$ & $\lambda_{\mathrm{C.M.}}$ & GSD\\
         \hline
         PH(111) & $0$ & $1$ & $1$ & $1$&$1$\\
         FCI  & max & $1$ & $1$ & $\{\sqrt[3]{1}\}$&$3=1\cdot3$\\
         PS I &  $0$& $(\pm1,\pm i)$& $1$&$1$&$4$\\
         PS II &  $0$& $(\pm1,\pm i)$ & $\pm1$ &$1$& $8$\\
         FTI &  $0$ & $1$ & $1$ & $1$ &$9$\\
         PS III &  $0$ & $(\pm1,\pm i)$ & $1$, $(\pm1)$ & $1$  &$56=2\cdot4+6\cdot8$         
    \end{tabular}
    \caption{Summary of the emergent phases on the TR symmetric LLL on the torus. The abbreviation PS stands for ``phase separated state". The values for the $\lambda_i$ columns represent the eigenvalues for the $C_4$, $y$-mirror ($M_y$) and center of mass ($\mathrm{C.M.}$) symmetries, respectively. The set of the three complex cube roots of $1$ are denoted by $\{\sqrt[3]{1}\}$. How the symmetry eigenvalues dictate the ground state degeneracy is discussed in appendix \ref{App1}. There, also representative spectra for the listed phases are provided. The last column lists the composition of the $\mathrm{GSD}$ in an abbreviated manner, where the $3$-fold redundancy from the center of mass symmetry in the case of full spin polarization and the decomposition into the distinct states for $\mathrm{GSD}=56$ are highlighted.}
    \label{tab:1}
\end{table}

Fig.~\ref{fig:3} shows the effect of individual pseudopotentials on the TR symmetric LLL in more detail by considering the dimension of the ground state manifold. The phase diagram on the torus, apart from gapless regions, can be divided into regions with different $\mathrm{GSD}$ and four families of phases arise. One of these families contains the spontaneously spin polarized phases discussed above. For even $m$, there is a stable spontaneously magnetized $\nu_\sigma=2/3$, $\nu_{\Bar{\sigma}}=0$ FCI as these even $m$ pseudopotentials do not affect the single spin species Laughlin state due to fermionic anti-symmetry.

Regions in the phase diagrams of relatively low $v_0$ exhibit a phase with a single ground state at $\bk=(0,0)$ and a comparatively large spectral gap. The dominant orbital occupation components of this state show double occupancy, indicating a paired state. Fig.~\ref{fig:4} shows that for $v_{m>1}=0$ the state is adiabatically connected to the phase at large negative $v_0$, which was shown to be an inter-spin paired $s$-wave superconductor of electrons by means of flux insertion \cite{crepel2024attractive}. Borrowing concepts from quantum Hall ferromagnetism, one can find an alternative description of this state. Particle-hole transforming the spin-down species in the relevant limit $v_0\rightarrow-\infty$ yields spins with equal Chern number $C_\uparrow=C_\downarrow=1$, filling fractions $\nu_\uparrow=1/3$, $\nu_\downarrow=2/3$ and interaction $v_0\rightarrow+\infty$. This has the Halperin (111) state \cite{Halperin:1983zz} as ground state. Particle-hole transforming one spin species to go back to the opposite Chern scenario we are considering, one obtains the partially particle-hole transformed Halperin (111) state. The state with $\mathrm{GSD}=1$ will therefore in the following be labeled PH(111).

The phases with $\mathrm{GSD} \mod 4=0$ have ground states that individually break the four-fold rotational symmetry. The multiplicity of $4$ is readily explained by having nonzero wave vectors for charge ordered phases and the orbit of this wave vector under the action of the symmetry group $C_4 \times M_y$, where $M_y$ is the mirror symmetry group associated with the $y$-axis. For spin polarized phase separated states, the $\mathrm{GSD}$ due to symmetry considerations is tripled. A more detailed discussion is given in appendix \ref{App1}. Spontaneous symmetry breaking is further indicated by the exact degeneracy of the ground states. The most prominent representative in the phase diagrams is the state of $\mathrm{GSD}=4$ and shows clear charge order upon investigation on the sphere (see appendix~\ref{App3}) and the wave vector of the charge order scales with system size. This is in agreement with the phase separated states found in Refs.~\onlinecite{Chen2012FQHtorusFTI,Mukherjee2019FQHsphereFTI}. The 8-fold degenerate state is likely gapless, and connected to the phase separated state of Ref.~\onlinecite{Chen2012FQHtorusFTI}. The 4-fold degenerate state is potentially gapless too, although further investigation is necessary. In the phase diagrams these phase separated states are abbreviated as $\mathrm{PS}$ I -- III, distinguished by their $\mathrm{GSD}$. The $\mathrm{PS}$ III phase is composed of $8$ individual phase separated states that are close in energy compared to the gap size and is therefore identified by their combined $\mathrm{GSD}$ of $2\cdot4 + 6\cdot 8 = 56$. The individual components are evident in Fig.~\ref{fig:app8}. 

Finally, the $\mathrm{GSD}=9$ phase is the FTI. For unscreened Coulomb-type interactions, the on-site interaction $v_0/v_1\approx2$ and the stabilization of the FTI therefore requires strong $v_0$ suppression \cite{kwan2024abelianfractionaltopologicalinsulators}. It is adiabatically connected to the limit of decoupled FCIs and survives isotropic inter-spin interactions at sufficiently low $v_0$ with respect to its Coulomb value, as seen from the figure in appendix \ref{App2}. Inter-spin coupling and non-zero $m\geq2$ pseudopotentials deteriorate the quality parameter and the FTI is unstable against stabilization of the PH(111) state or phase separated phases. The ideal value of $v_0/v_1$ for the FTI drifts as a function of $v_{m\geq2}$. The trend with $v_{m\geq2}$ follows an even-odd pattern in $m$, according to the boundaries of the PH(111) phase. The identified phases can be summarized in a table according to the respective $\mathrm{GSD}$ and spin polarization (Tab.~\ref{tab:1}).\\


\section{Conclusion}


In this work, we study how the individual Haldane pseudopotentials affect the FTI in TR symmetric lowest Landau levels. The different regions in the $(v_m,v_0)$ phase diagrams have been identified and classified in order to understand the competing phases to the FTI. We found the partially particle-hole transformed Halperin (111) state which is adiabatically connected to a superconductor, a family of polarized and unpolarized phase separated states, a spin polarized FCI and the sought-after FTI. It is evident that there is an even-odd pattern in the relative angular momentum $m$, owed to the effect of $v_m$ as an inter-spin or intra-spin interaction. Large non-zero $v_{m\geq3}$ generally destroys the FTI, however, for small interaction amplitudes the FTI can be stabilized for appropriate $v_0$. For a pure Coulomb interaction, $v_0/v_1 \approx 2$, such that it has to be diminished drastically in order to stabilize the FTI at $v_0/v_1\sim1.25$. This condition can be relaxed if small long-range interactions are included. To achieve this suppression and access the relevant regions of the phase diagram in terms of the FTI, an additional short-range interaction has to be introduced. One way to accomplish such an additional background potential is by dielectric engineering \cite{kwan2024abelianfractionaltopologicalinsulators}. 


\begin{acknowledgements}
We thank Prof.~Manfred Sigrist for insightful discussions. G.W.~is supported by the Swiss National Science Foundation (SNSF) via Ambizione grant number PZ00P2-216183. Exact diagonalization calculations were performed using DiagHam. T.N.~acknowledges support from the Swiss National Science Foundation through a Consolidator Grant (iTQC, TMCG-2\_213805). 
\end{acknowledgements}

\bibliography{refs}

\newpage
\clearpage
\begin{appendix}
\onecolumngrid
	\begin{center}
		\textbf{\large --- Supplementary Material ---\\Tuning between a fractional topological insulator and competing phases at $\nu_\mathrm{T}=2/3$}\\
		\medskip
		\text{Roger Brunner, Titus Neupert, Glenn Wagner}
	\end{center}
	
		\setcounter{equation}{0}
	\setcounter{figure}{0}
	\setcounter{table}{0}
	\setcounter{page}{1}
	\makeatletter
	\renewcommand{\theequation}{S\arabic{equation}}
	\renewcommand{\thefigure}{S\arabic{figure}}
	\renewcommand{\bibnumfmt}[1]{[S#1]}

\section{Full set of phase diagrams}
\label{App0}

\begin{figure}[htpb!]
    \centering
    \includegraphics[width = \linewidth]{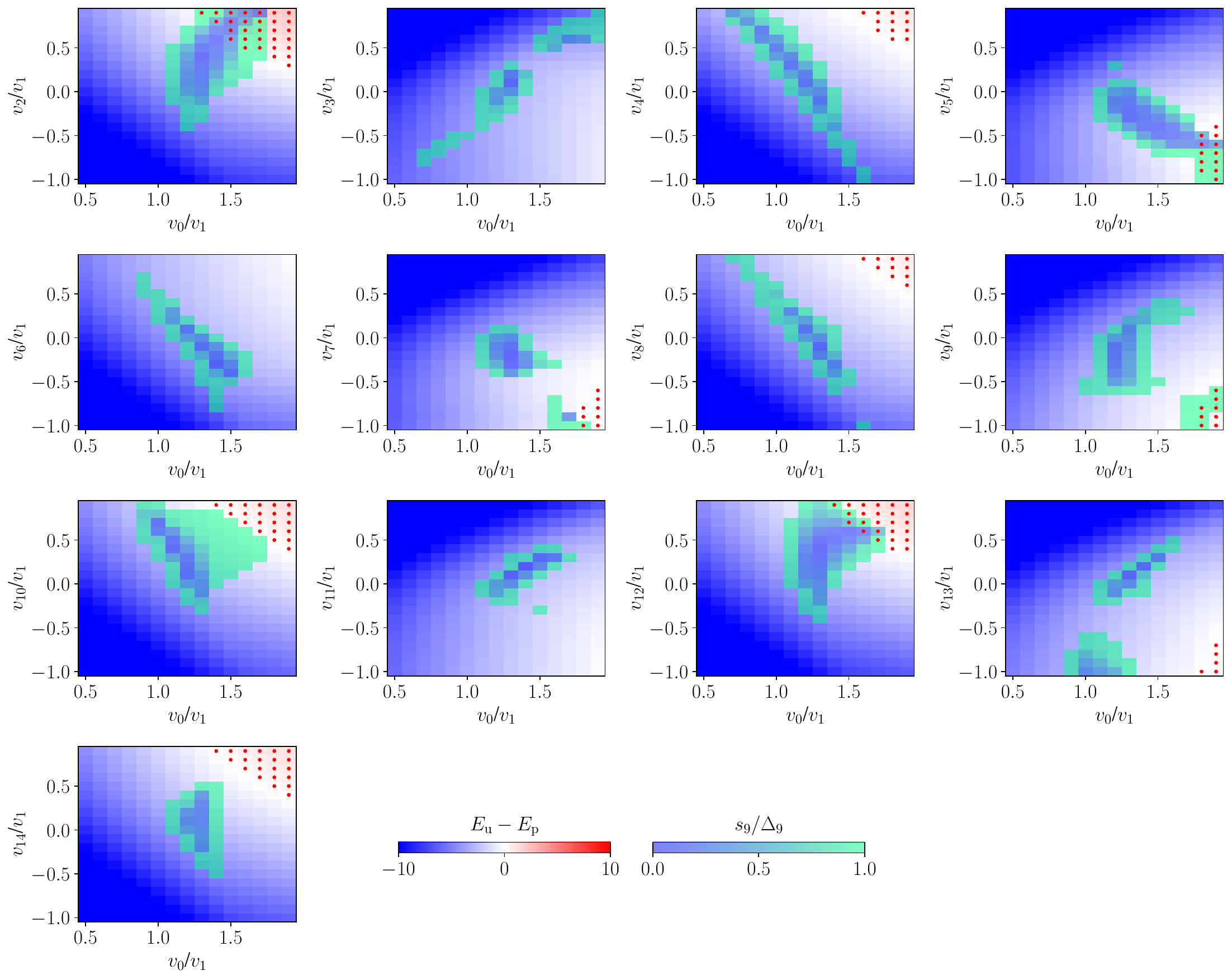}
    \caption{Phase diagrams for the ground state spin polarization as a function of $v_m$ and $v_0$ (for fixed $v_1$) measured by the difference between the ground state energies of the spin unpolarized and spin polarized sectors, $E_\mathrm{u}-E_\mathrm{p}$. The system parameters are $N_\Phi=15$ and $N_e=10$. The red dots mark spin polarized regions. There, there are FCIs as exact ground states for $v_{2n}$, FCIs in competition with phase separated states for some $v_{2n+1}$ and purely phase separated states for others. The FTI quality parameter $0\leq s_9/\Delta_9 < 1$ (green-blue regions) is shown to illustrate its alignment with equi-energy-difference lines, which is evident for all even $m$. Competition with full spin polarization appears to be beneficial to the FTI. The apparent even-odd effect with respect to the favoured polarization at repulsive or attractive interactions is owed to the fact that $v_{2n}$ are purely inter-spin interactions and hence the energy penalty for repulsive interactions can be avoided by polarization. For $v_{2n+1}$, which include intra-spin interactions, polarization is favoured for attractive interactions.}
    \label{fig:Appspinpol}
\end{figure}

\pagebreak

\begin{figure}[htpb!]
    \centering
    \includegraphics[width = \linewidth]{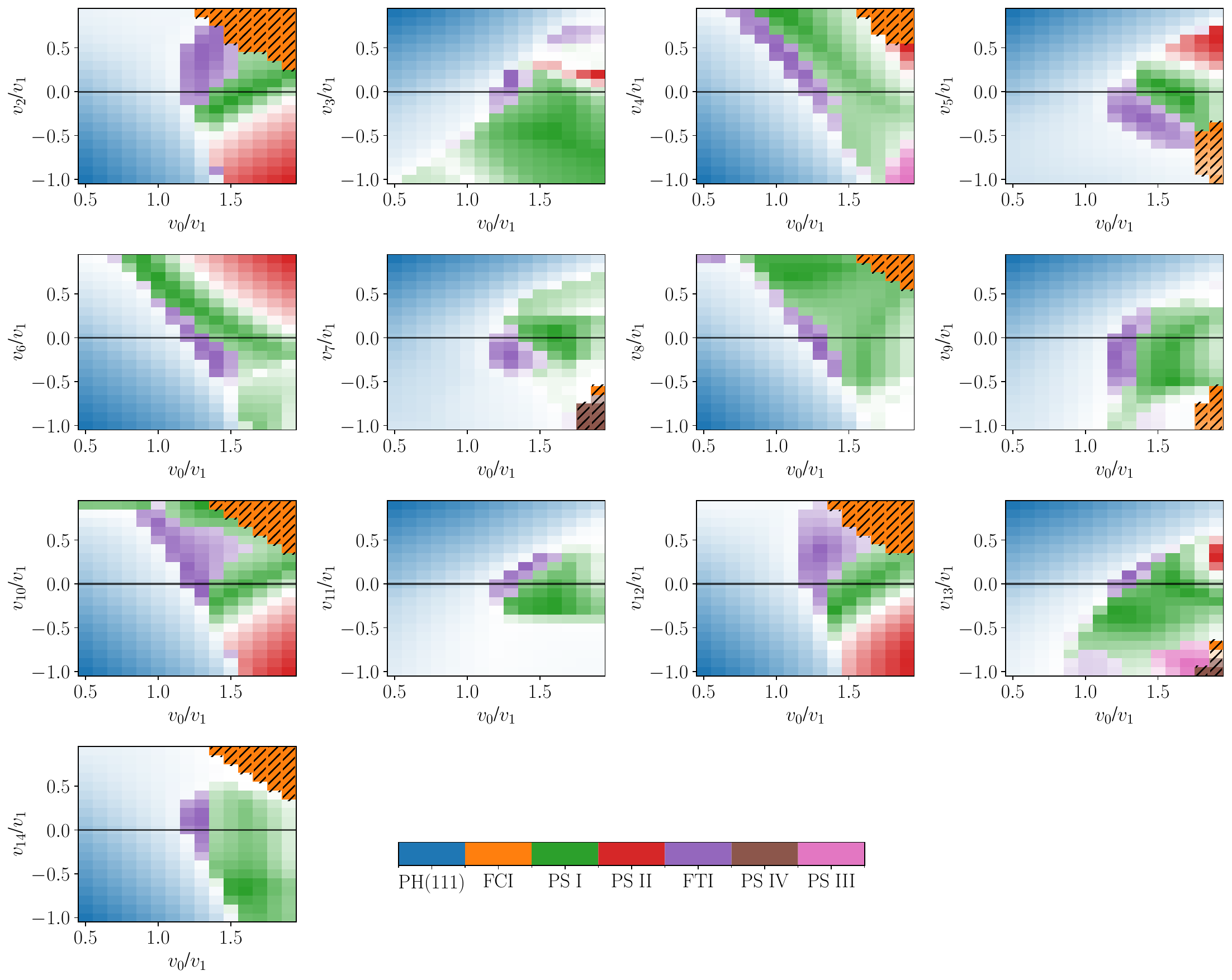}
    \caption{Phase diagrams for the ground state degeneracy in terms of individually tuned $v_m$ and $v_0$ for fixed $v_1$. The system parameters are $N_\Phi=15$ and $N_e=10$. The colors determine the phases according to their respective $\mathrm{GSD}$ and the transparency reflects the size of the quality parameter $\Delta_{\mathrm{GSD}}/\max_{\{(v_0,v_m)\}}\Delta_\mathrm{GSD}$, being fully opaque when it is one. Hatched regions indicate full spin polarization and white regions reflect spectra for which a $\mathrm{GSD}$ cannot be identified conclusively or are gapless. Featured in the phase diagrams are the partially particle-hole transformed Halperin (111) state ($\mathrm{GSD}=1$), the FCI in the polarized region ($\mathrm{GSD}=3$), the FTI ($\mathrm{GSD}=9$) and polarized and unpolarized phase separated states, i.e. $\mathrm{PS}$ I-IV, of different ground state momentum sectors ($\mathrm{GSD}\mod 4 =0$). $\mathrm{PS}$ IV is not featured in the phase diagrams in the main text. The details of all phases are summarized in table \ref{tab:app1}. To discriminate against false detection of ground state degeneracies during transitions, the condition of featuring the correct momentum sectors was added. There is an evident even-odd effect in $m$, which dictates the direction and extent of the FTI phase in the phase diagram. Spectra with the associated ground state momentum sectors for the individual phases are illustrated in appendix \ref{App1}.}
    \label{fig:Appphasediag}
\end{figure}

\begin{table}[htpb!]
    \centering
    \begin{tabular}{c|c|c||c|c|c||c}
          $\mathrm{GSD}$ & Phase  & Spin Pol. & $\lambda_{C_4}$ & $\lambda_{M_y}$ & $\lambda_{\mathrm{C.M.}}$ & deg. from sym.\\
         \hline
         $1$ & PH(111) & $0$ & $1$ & $1$ & $1$&$1$\\
         $3$ & FCI  & max & $1$ & $1$ & $\{\sqrt[3]{1}\}$&$3\cdot1$\\
         $4$ & PS I &  $0$& $(\pm1,\pm i)$& $1$&$1$&$4$\\
         $8$ & PS II &  $0$& $(\pm1,\pm i)$ & $\pm1$ &$1$& $8$\\
         $12$ & PS IV & max & $(\pm1,\pm i)$ & 1 & $\{\sqrt[3]{1}\}$ & $3\cdot4$ \\
         $9$ & FTI &  $0$ & $1$ & $1$ & $1$ &$1$\\
         $56$ & PS III &  $0$ & $(\pm1,\pm i)$ & $1$, $(\pm1)$ & $1$ &$2\cdot4$ + $6\cdot8$         
    \end{tabular}
    \caption{Summary of the emergent phases on the TR symmetric LLL on the torus. The abbreviation PS stands for ``phase separated state". The values for the $\lambda_i$ columns represent the eigenvalues for the $C_4$, $y$-mirror ($M_y$) and center of mass ($\mathrm{C.M.}$) symmetries, respectively. The set of the three complex cube roots of $1$ are denoted by $\{\sqrt[3]{1}\}$. How the symmetry eigenvalues dictate the ground state degeneracy is discussed in appendix \ref{App1}. There, also representative spectra for the listed phases are provided. The last column lists the composition of the $\mathrm{GSD}$ in an abbreviated manner, where the $3$-fold redundancy from the center of mass symmetry in the case of full spin polarization and the decomposition into the distinct states for $\mathrm{GSD}=56$ are highlighted.}
    \label{tab:app1}
\end{table}

\pagebreak

\section{Representative spectra for the listed phases}
\label{App1}

\begin{figure}[htbp!]
    \centering
    \subfigure[]{\includegraphics[width = 0.32 \linewidth]{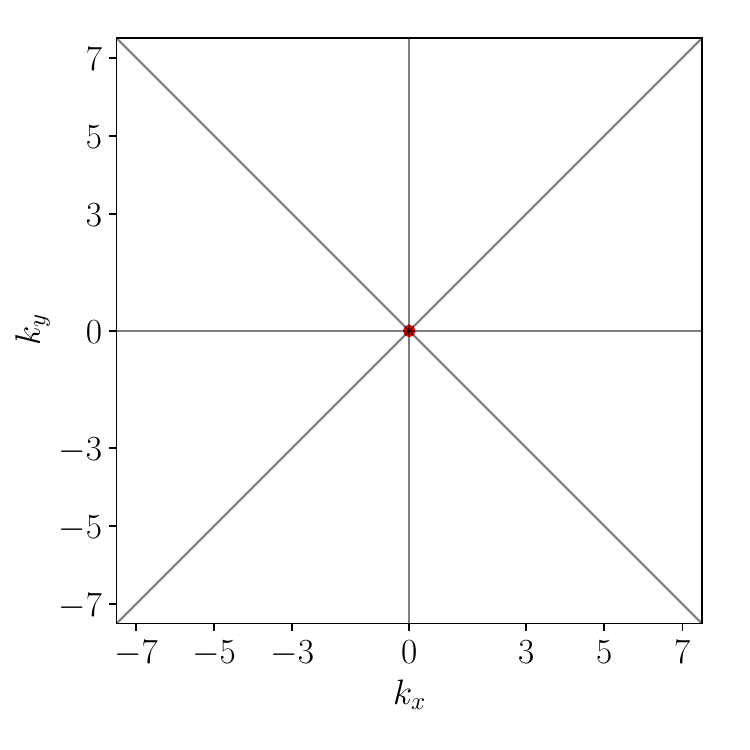}}
    \subfigure[]{\includegraphics[width = 0.32 \linewidth]{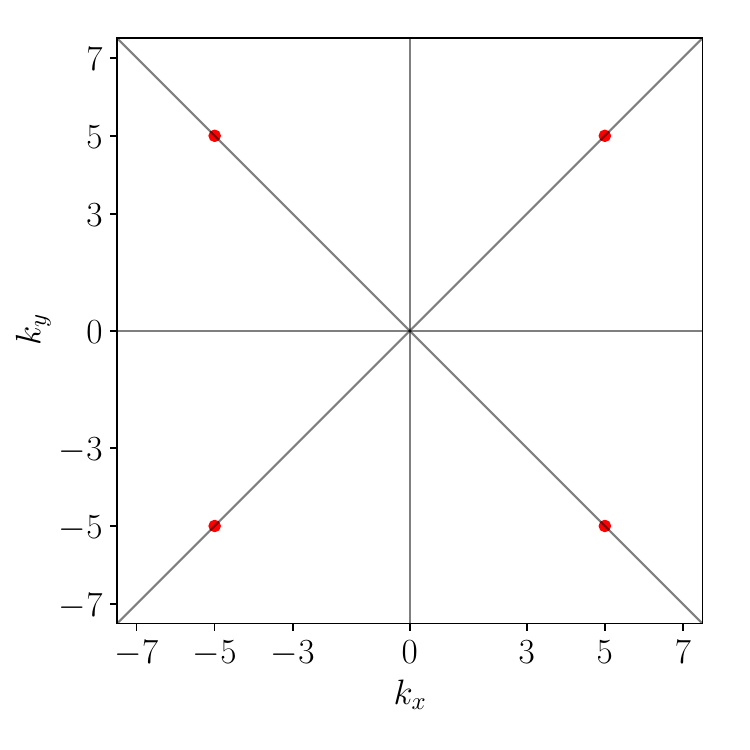}}
    \subfigure[]{\includegraphics[width = 0.32 \linewidth]{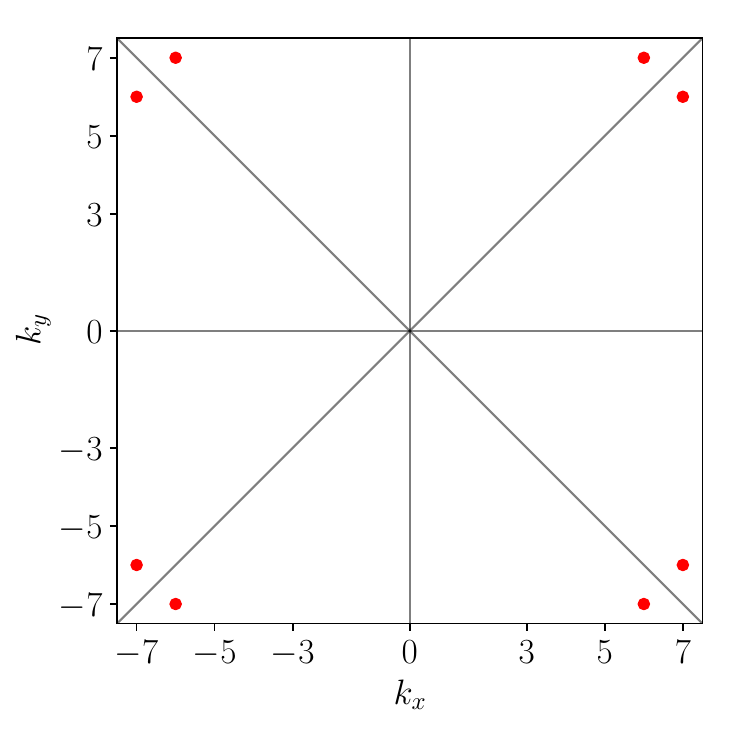}}
    \hfill
    \subfigure[]{\includegraphics[width = 0.32 \linewidth]{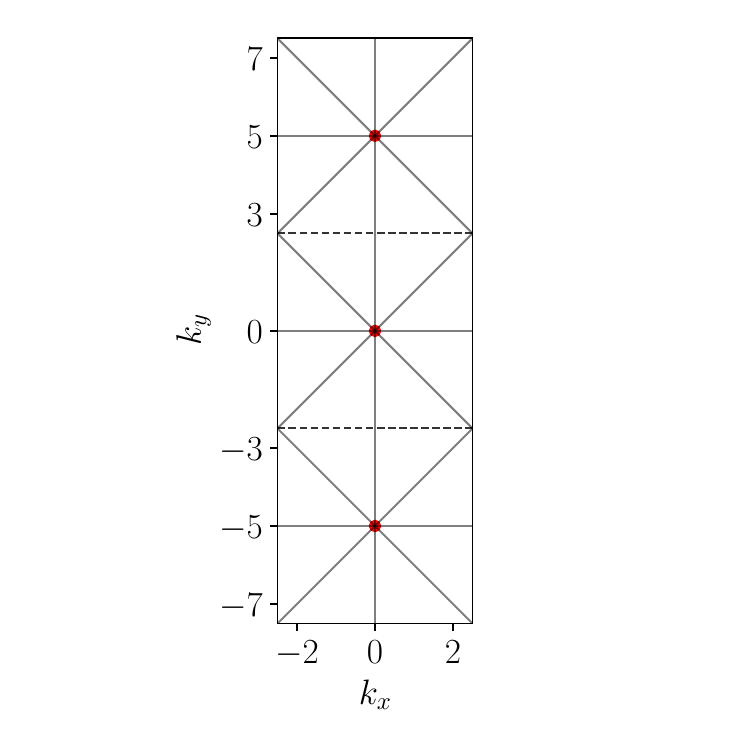}}
    \subfigure[]{\includegraphics[width = 0.32 \linewidth]{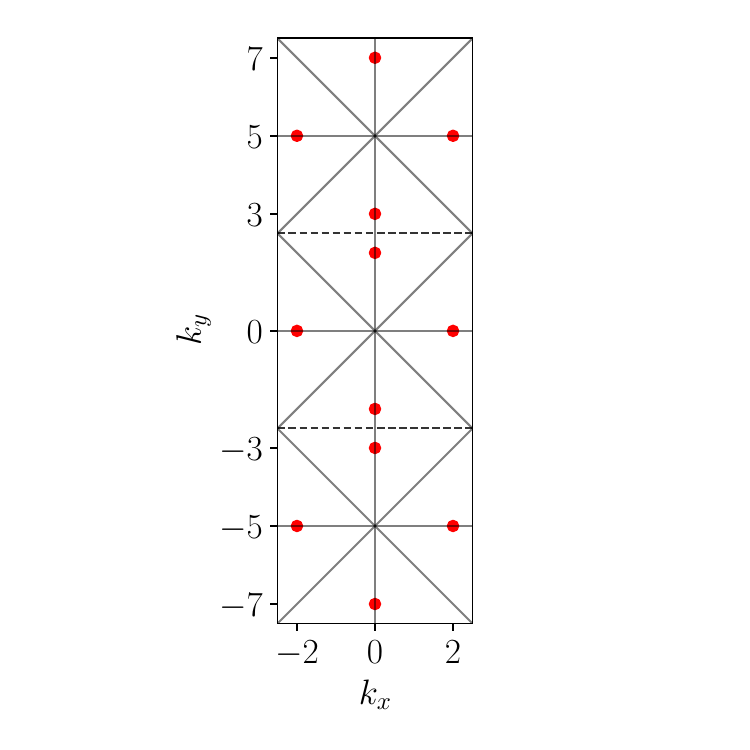}}
    \subfigure[]{\includegraphics[width = 0.32 \linewidth]{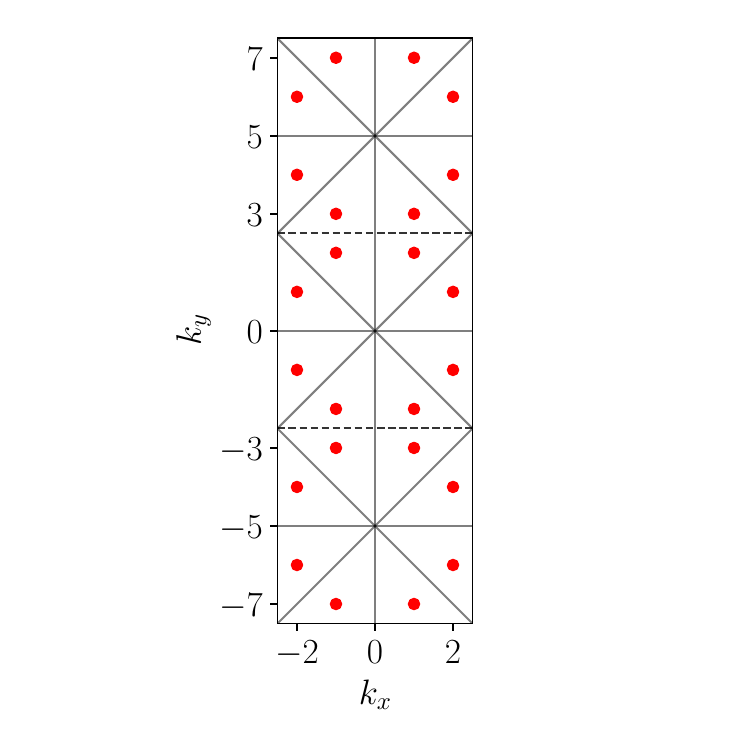}}
    \caption{(a) - (c) Visualization of the different momentum sector sizes for the case of zero net spin polarization. There is a single state for the sector $(0,0)$, $4$ states for sectors that lie along high-symmetry axes and $8$ states for any other points in the many body momentum space. (d) - (f) Cases exhibiting the same effect, but for maximal spin polarization. The tripling of the reduced momentum sectors mentioned in the text is identifiable along the $k_y$ axis and will be omitted in the following momentum sector plots. The identification $k_i \cong k_i-N$, where $N=15$ for the unpolarized and $N=5$ for the polarized case (in $k_x$-direction), is due to periodic boundary conditions in both directions on the torus.}
    \label{symm}
\end{figure}

\noindent The multiplicity of the degenerate ground states for the phase separated phases can be explained in terms of the symmetries of the system and the many body momentum sector of the ground states. The symmetry group is $G=C_4 \times M_y$, where $M_y$ is the mirror symmetry group that acts on a point in momentum space as $M_y(k_x,k_y)=(k_x,-k_y)$. This symmetry is implied by TR symmetry and the other mirror symmetry group $M_x$ is a subgroup of $G$. If a ground state momentum $\bk$ lies in a low-symmetry point, the orbit $G_{\bk}$ contains $8$ points. For points along the high-symmetry axes $k_i=0$ and $k_x = \pm k_y$ the orbit has cardinality $4$, and contains only a single point for the origin $\bk=(0,0)$. Examples are given in Figs.~\ref{symm} (a)-(c). The momentum sector of the Halperin (111) state is in the origin and, thus, nondegenerate. Phase separated states necessarily exhibit a non-zero wave vector and therefore feature an intrinsic degeneracy of $4$ or $8$.

Full spin polarization reduces the momentum space by a factor of 9, according to the discussion in Ref.~\onlinecite{Haldane1985manybody}. Drawing from the notation of the above reference, the studied system features $N_e=p\cdot N=2\cdot5$ and $N_\Phi=q\cdot N = 3\cdot5$, where $N$ is the largest common divider of $N_e$ and $N_\Phi$, such that the momentum space contains $N^2 = 25$ instead of $225$ points and exhibits an intrinsic degeneracy of $q=3$ by the center of mass translation symmetry. For the sake of identification, this degeneracy is reintroduced in the spectrum plots and phase diagrams, as to e.g. feature the $3$-fold ground state degeneracy of the Laughlin state on the torus. This will be reflected in the tripling of the reduced $k_y$-axis, such that $k_x \in \{0,\dots,5\}$ and $k_y\in\{0,\dots,15\}$. For this reason, the degeneracies caused by the symmetry group $G$ are therefore featured $3$-fold for the spin polarized case. Examples thereof are shown in Figs.~\ref{symm} (d)-(f).

\newpage

\pagebreak

\begin{figure}[htpb!]
    \centering
    \includegraphics[width=0.45\linewidth]{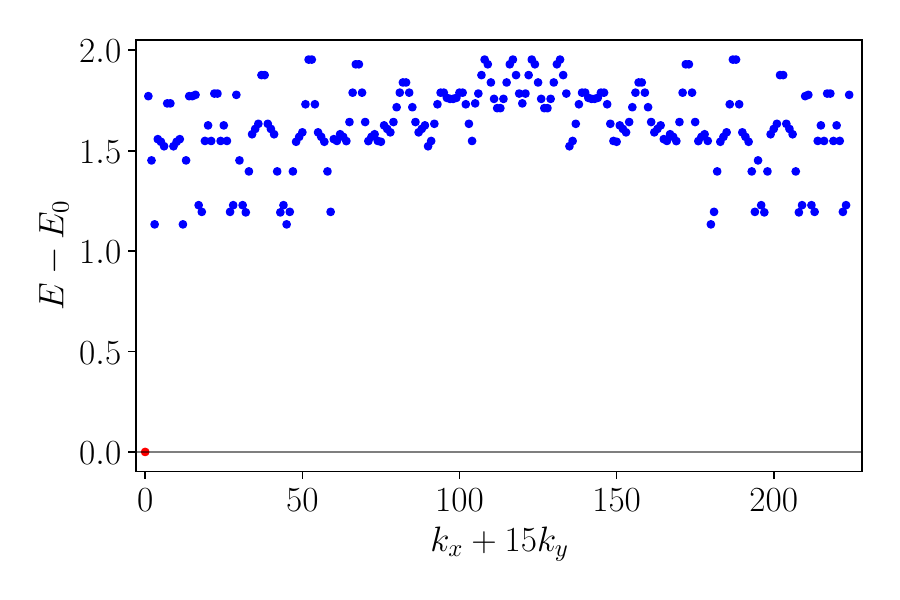}
    \caption{Representative spectrum of the phase of $\mathrm{GSD}=1$, i.e. the PH(111) state. The system parameters are $N_\Phi=15$, $N_e=10$ at $v_0/v_1 = 0.5$ and $v_3/v_1=0$. The single ground state at $\bk=(0,0)$ and the relatively large spectral gap are clearly identifiable.}
    \label{fig:app1}
\end{figure}

\begin{figure}[htpb!]
    \centering
    \includegraphics[width=0.45\linewidth]{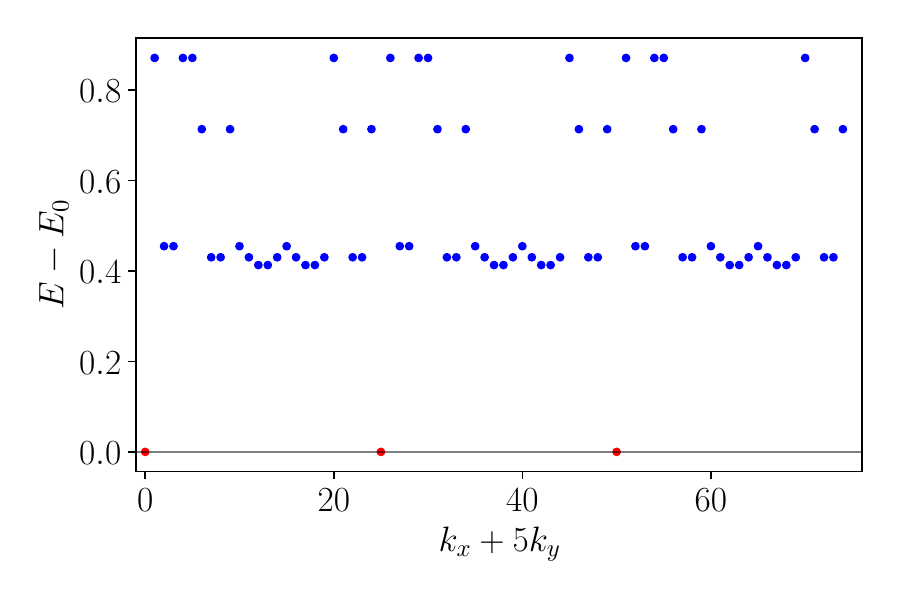}
    \caption{Representative spectrum of the phase of $\mathrm{GSD}=3$ at maximal spin polarization, i.e. the FCI. The system parameters are $N_\Phi=15$, $N_e=10$ at $v_0/v_1 = 1.7$ and $v_2/v_1=0.7$. As $m$ is even, the ground state is the Laughlin state $\ket{\frac{1}{3}}\otimes {\ket{\bar{\frac{1}{3}}}}$ and the system thus an ideal FCI. The momentum space is reduced by a factor of $3$ in the $k_x$ direction according to the details in the text above.}
    \label{fig:app2}
\end{figure}

\pagebreak

\begin{figure}[htbp!]
    \centering
    \subfigure[]{\includegraphics[width = 0.45 \linewidth]{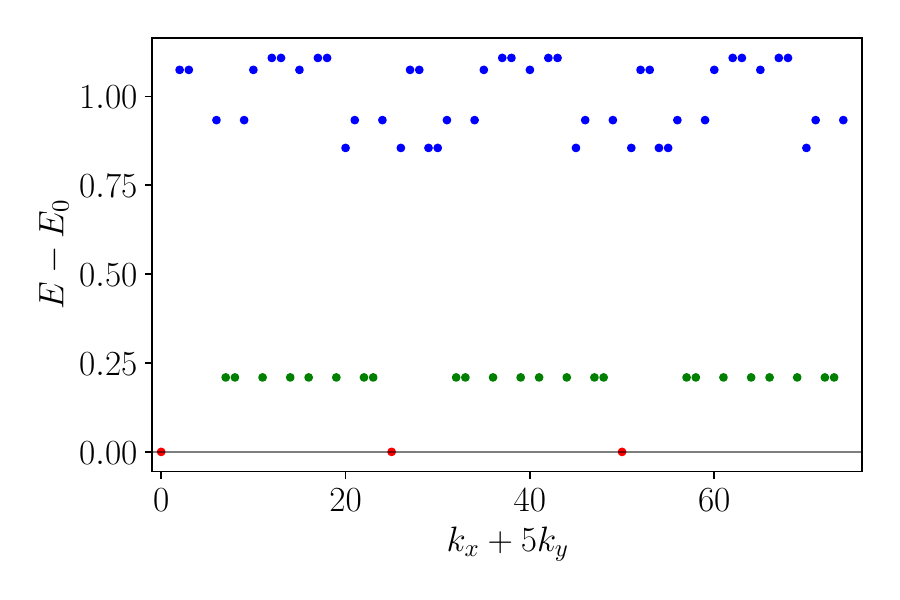}}
    \subfigure[]{\includegraphics[width = 0.45 \linewidth]{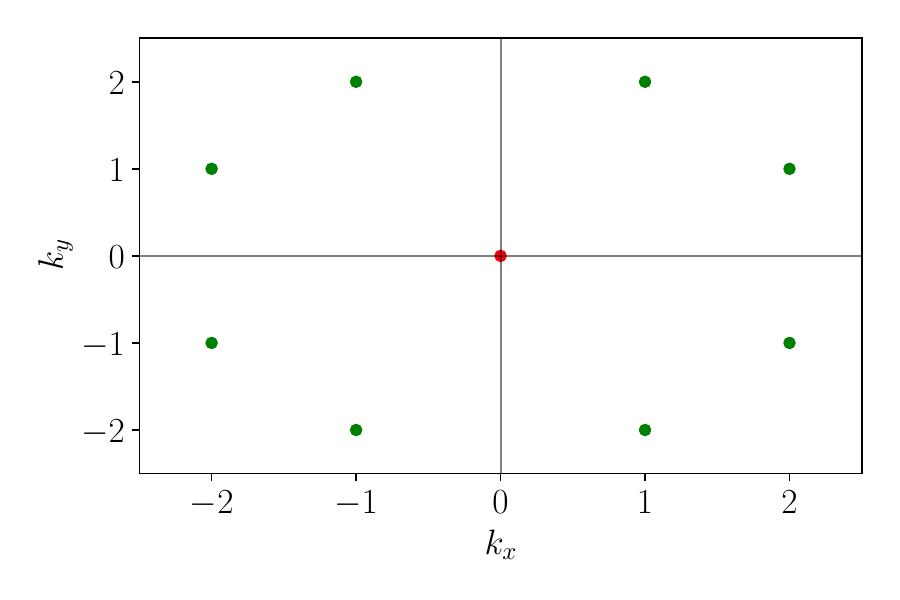}}
    \caption{(a) Another representative spectrum of the phase of $\mathrm{GSD}=3$ at maximal spin polarization, i.e. the FCI. The system parameters are $N_\Phi=15$, $N_e=10$ at $v_0/v_1 = 1.9$ and $v_5/v_1=-1$. As $m$ is odd, the Laughlin state is the ground state at $v_5=0$ but deviates with increasing $v_5$. In the probed region, the FCI competes with a phase separated phase of $\mathrm{GSD}=3\cdot8=24$ but remains in the ground state. The momentum space is reduced by a factor of $3$ in the $k_x$ direction according to the details in the text above. (b) Visualization of the ground state momentum sector. The FCI ground state lies at $\bk=(0,0)$. Points in the momentum sector of the competing phase separated state do not lie on any high-symmetry axis and the resulting number of states in the momentum sector is therefore $8$.}
    \label{fig:app2a}
\end{figure}

\begin{figure}[htbp!]
    \centering
    \subfigure[]{\includegraphics[width = 0.45 \linewidth]{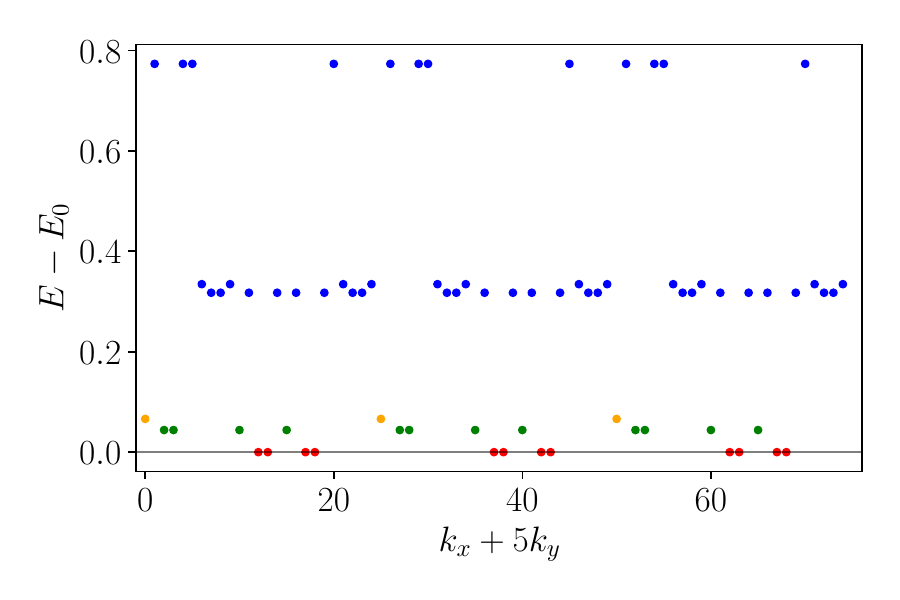}}
    \subfigure[]{\includegraphics[width = 0.45 \linewidth]{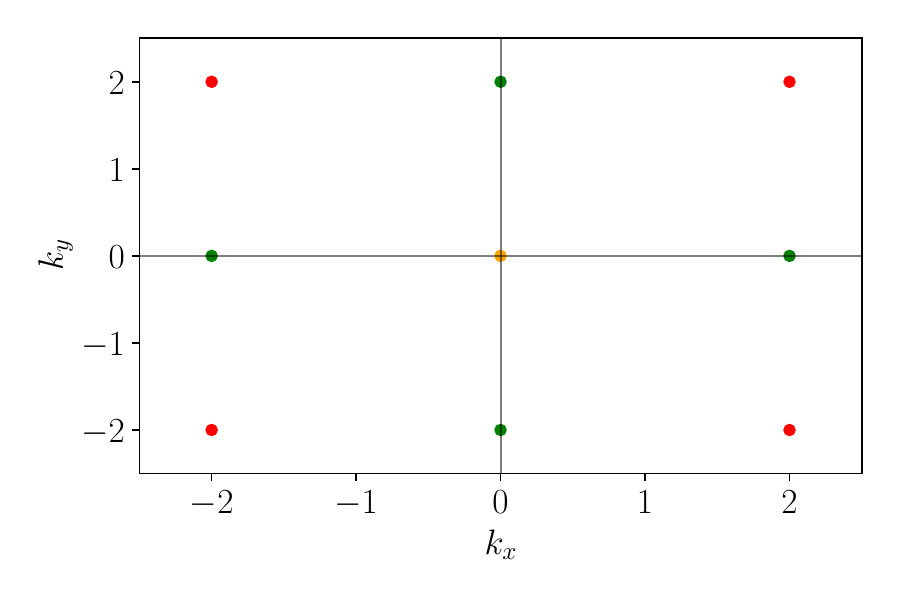}}
    \caption{(a) Spectrum in the context of Figs.~\ref{fig:app2} and \ref{fig:app2a} for maximal spin polarization, where the ground state transitions away from the FCI into a phase separated state. The system parameters are $N_\Phi=15$, $N_e=10$ at $v_0/v_1 = 1.9$ and $v_{13}/v_1=-1$. As $m$ is odd, the Laughlin state is the ground state at $v_{13}=0$ but deviates with increasing $v_{13}$. In the probed region, the FCI competes with two phase separated phases of $\mathrm{GSD}=3\cdot4=12$ and gives way to one one of these phases upon decreasing $v_{13}$. The momentum space is reduced by a factor of $3$ in the $k_x$ direction according to the details in the text above. (b) Visualization of the ground state momentum sector. The FCI ground state lies at $\bk=(0,0)$. Points in the momentum sector of the competing phase separated states lie on high-symmetry axes and the resulting number of states in the momentum sectors is therefore $4$ each.}
    \label{fig:app2b}
\end{figure}

\pagebreak

\begin{figure}[htpb!]
    \centering
    \includegraphics[width=0.45\linewidth]{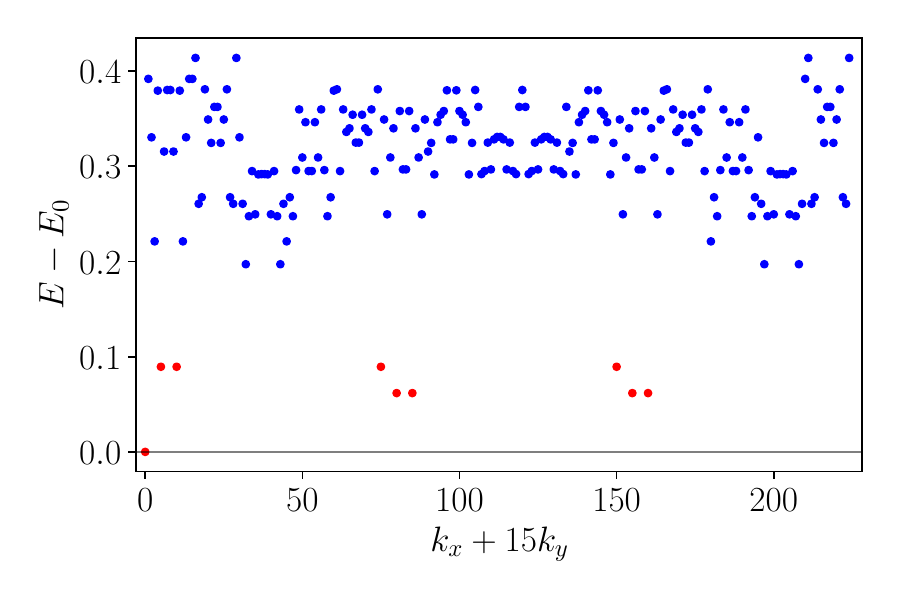}
    \caption{Representative spectrum of the phase of $\mathrm{GSD}=9$, i.e. the FTI. The system parameters are $N_\Phi=15$, $N_e=10$ at $v_0/v_1 = 1.2$ and $v_3/v_1=0$. The $9$ ground states in the topological momentum sector are visible with the lifted degeneracy due to inter-spin correlations.}
    \label{fig:app3}
\end{figure}

\begin{figure}[htbp!]
    \centering
    \subfigure[]{\includegraphics[width = 0.45 \linewidth]{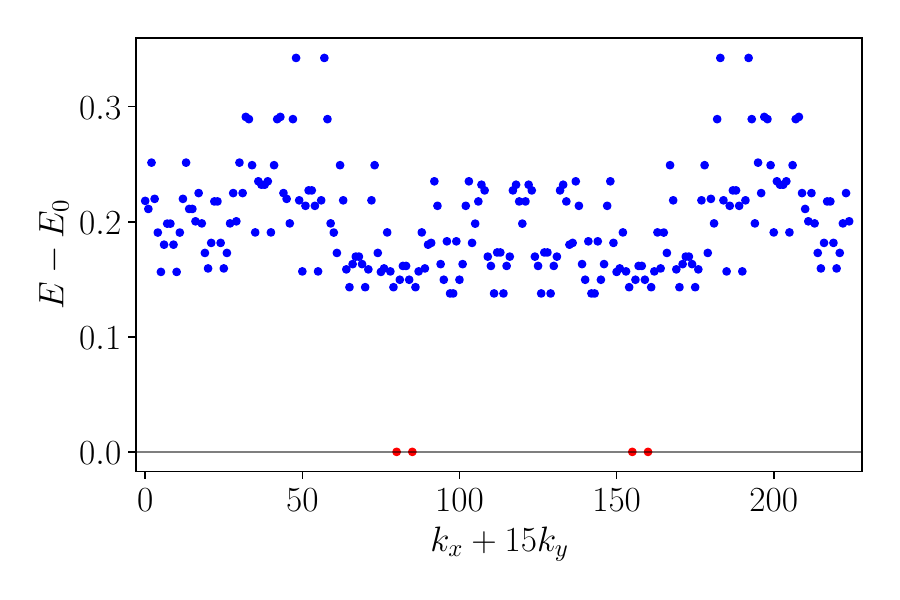}}
    \subfigure[]{\includegraphics[width = 0.45 \linewidth]{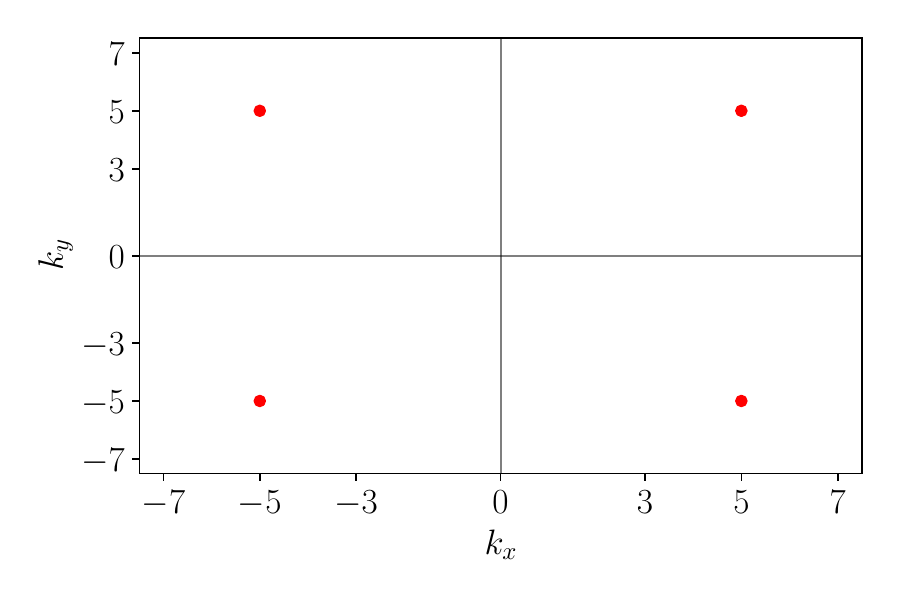}}
    \caption{(a) Representative spectrum of the phase of $\mathrm{GSD}=4$, i.e. the phase separated state I. The system parameters are $N_\Phi=15$, $N_e=10$ at $v_0/v_1 = 1.6$ and $v_3/v_1=-0.5$. $4$ degenerate ground states from spontaneous symmetry breaking are evident. (b) Visualization of the ground state momentum sector. The points in the momentum sector of this state lie on the high-symmetry axis $k_x=\pm k_y$ and the resulting number of states in the momentum sector is therefore $4$.}
    \label{fig:app4}
\end{figure}

\pagebreak

\begin{figure}[htbp!]
    \centering
    \subfigure[]{\includegraphics[width = 0.45 \linewidth]{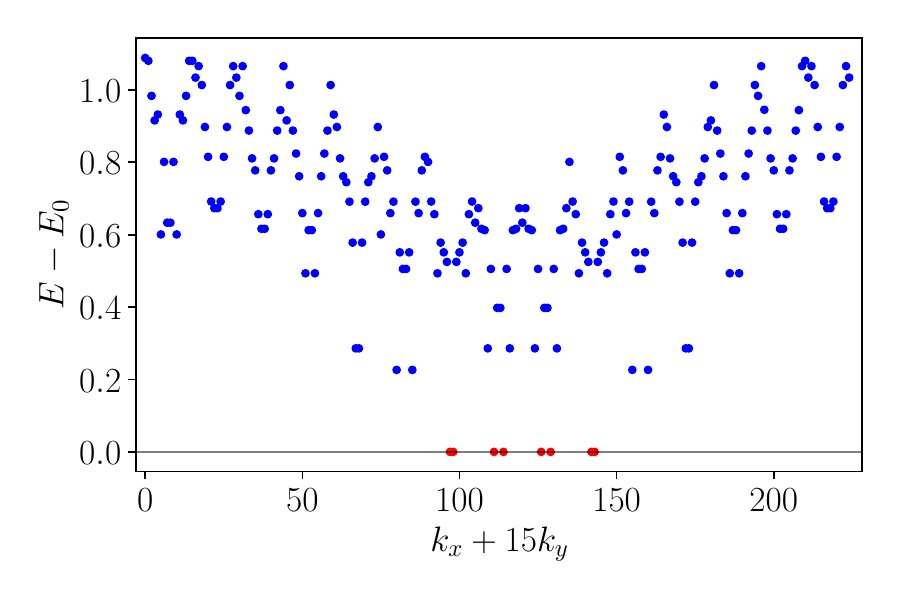}}
    \subfigure[]{\includegraphics[width = 0.45 \linewidth]{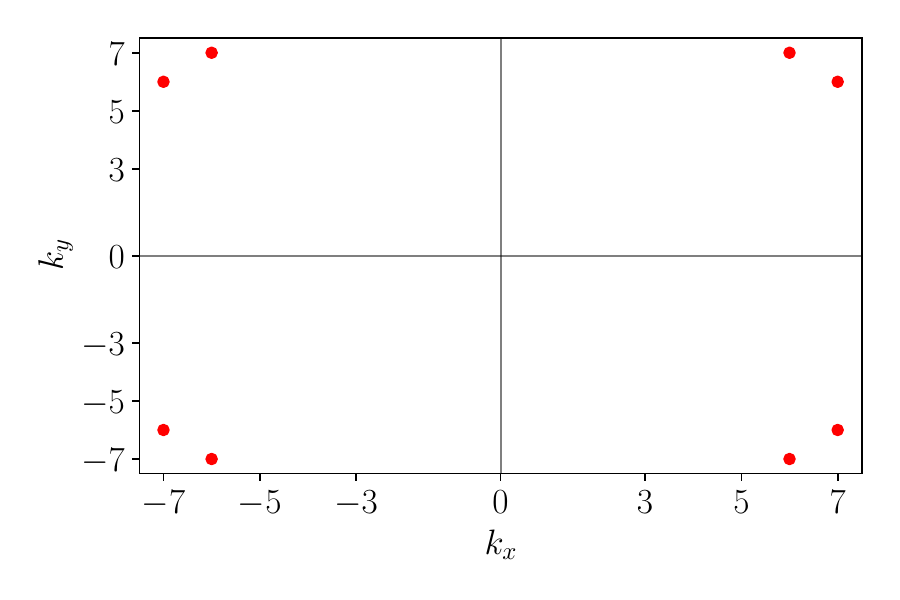}}
    \caption{(a) Representative spectrum of the phase of $\mathrm{GSD}=8$, i.e. the phase separated state II. The system parameters are $N_\Phi=15$, $N_e=10$ at $v_0/v_1 = 1.9$ and $v_6/v_1=0.9$. $8$ degenerate ground states from spontaneous symmetry breaking are evident. This phase visually resembles a gapless phase as the gap is small in the context of the full spectrum and the excited spectrum is not flat. This property largely remains until transitioning into the PH(111) phase upon increasing $v_6$. (b) Visualization of the ground state momentum sector. The points in the momentum sector of this state do not lie on any high-symmetry axis and the resulting number of states in the momentum sector is therefore $8$.}
    \label{fig:app5}
\end{figure}

\begin{figure}[htbp!]
    \centering
    \subfigure[]{\includegraphics[width =0.45 \linewidth]{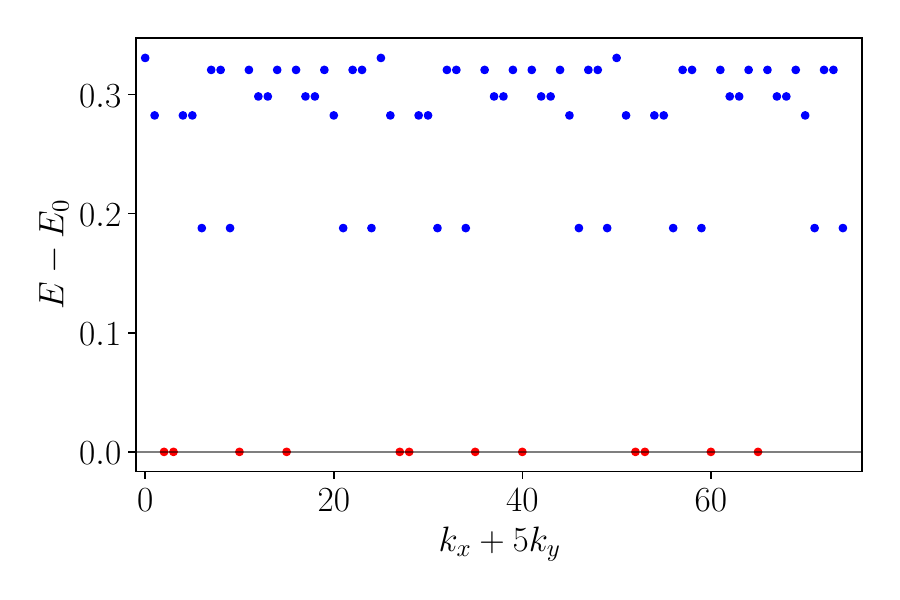}}
    \subfigure[]{\includegraphics[width = 0.45 \linewidth]{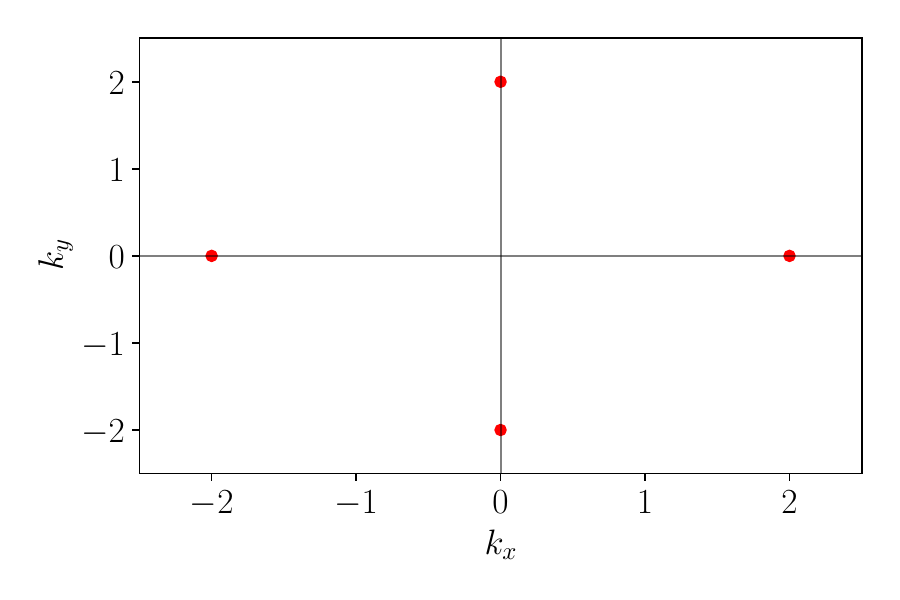}}
    \caption{(a) Representative spectrum of the phase of $\mathrm{GSD}=3\cdot4=12$ at maximal spin polarization, i.e. the phase separated state IV. The system parameters are $N_\Phi=15$, $N_e=10$ at $v_0/v_1 = 1.9$ and $v_7/v_1=-1.0$. The momentum space is reduced by a factor of $3$ in the $k_x$ direction according to the details in the text above. $3$ copies of $4$ degenerate ground states from spontaneous symmetry breaking are evident. (b) Visualization of the ground state momentum sector. The points in the momentum sector of this state lie on the high-symmetry axes $k_i =0$ and the resulting number of states in the reduced momentum sector is therefore $4$.}
    \label{fig:app6}
\end{figure}

\pagebreak

\begin{figure}[htbp!]
    \centering
    \subfigure[]{\includegraphics[width = 0.45 \linewidth]{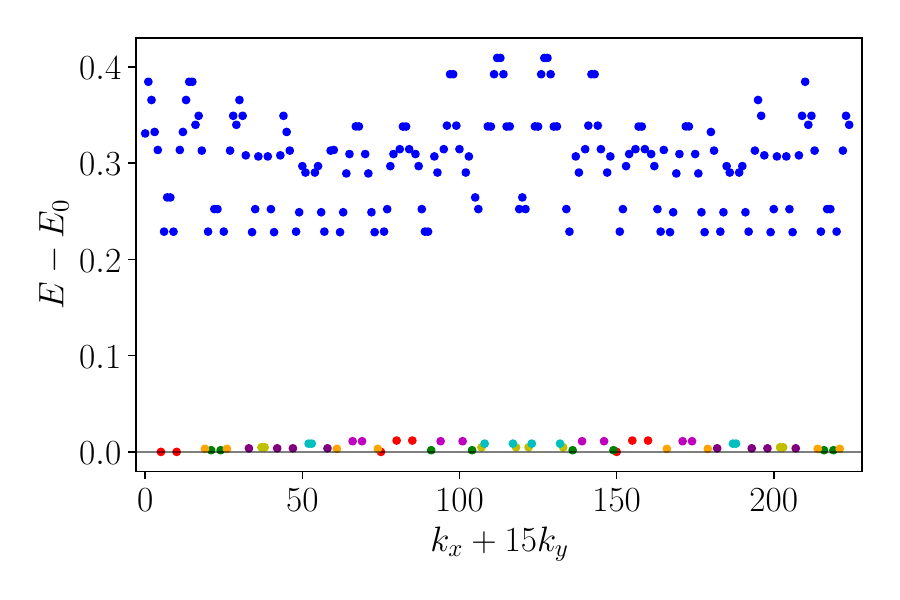}}
    \subfigure[]{\includegraphics[width = 0.45 \linewidth]{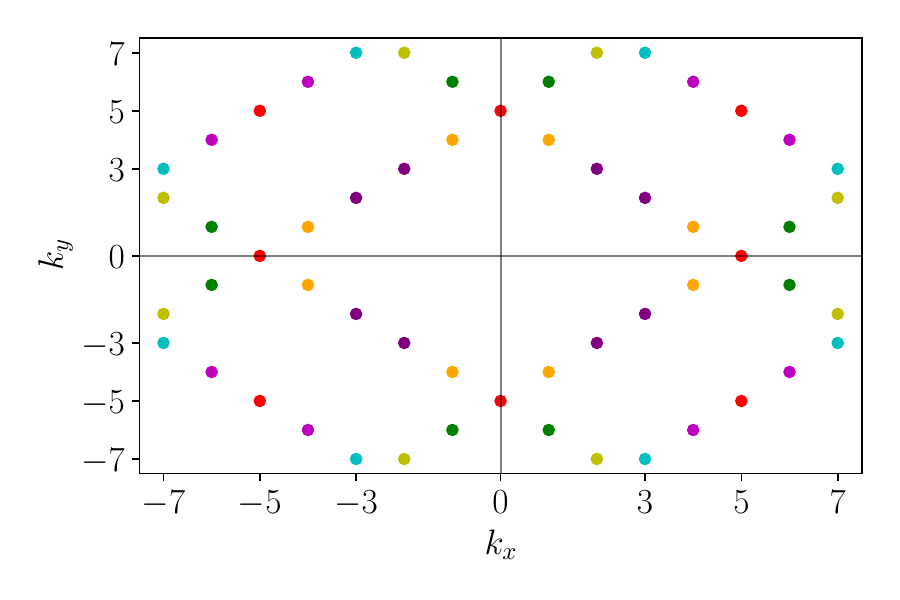}}
    \subfigure[]{\includegraphics[width = 0.45 \linewidth]{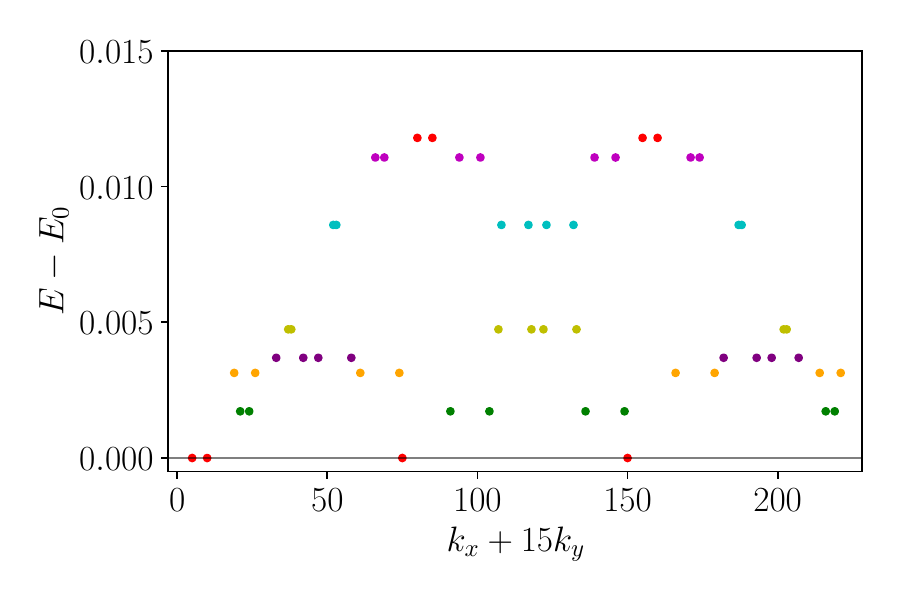}}
    \caption{(a) Representative spectrum of the phase of $\mathrm{GSD}=2\cdot 4 +6 \cdot 8 =56$, i.e. the phase separated state III. The system parameters are $N_\Phi=15$, $N_e=10$ at $v_0/v_1 = 1.9$ and $v_4/v_1=-0.9$. The $8$ individual phase separated states combine into a $56$-fold almost degenerate ground state, which, next to the large gap, is why it was detected as such. $2$ phase separated phases in the extended ground state exhibit momentum sectors in high-symmetry axes with associated degeneracy of $4$, whereas the degeneracy for the $6$ other phase separated phases in the extended ground state is $8$-fold. (b) Visualization of the ground state momentum sectors. There are 2 orbits of cardinality $4$ and $6$ orbits of cardinality $8$. (c) Spectrum with reduced $y$-axis for better distinguishability of the $8$ states in the ground state region. Exact degeneracy within the states of a given phase is evident.}
    \label{fig:app8}
\end{figure}

\pagebreak

\section{Connection to the decoupled limit on the torus}
\label{App2}

\noindent In order to discuss gradual inter-spin coupling, we introduce the inter-spin coupling parameter $\lambda\in[0,1]$ to the second term in the Hamiltonian \ref{eq: Hamil}. $\lambda=0$ corresponds to the limit where the two spin species are completely decoupled and $\lambda=1$ corresponds to isotropic inter-spin coupling, which is also the physical regime.

\begin{figure}[htpb!]
    \centering
    \includegraphics[width=\linewidth]{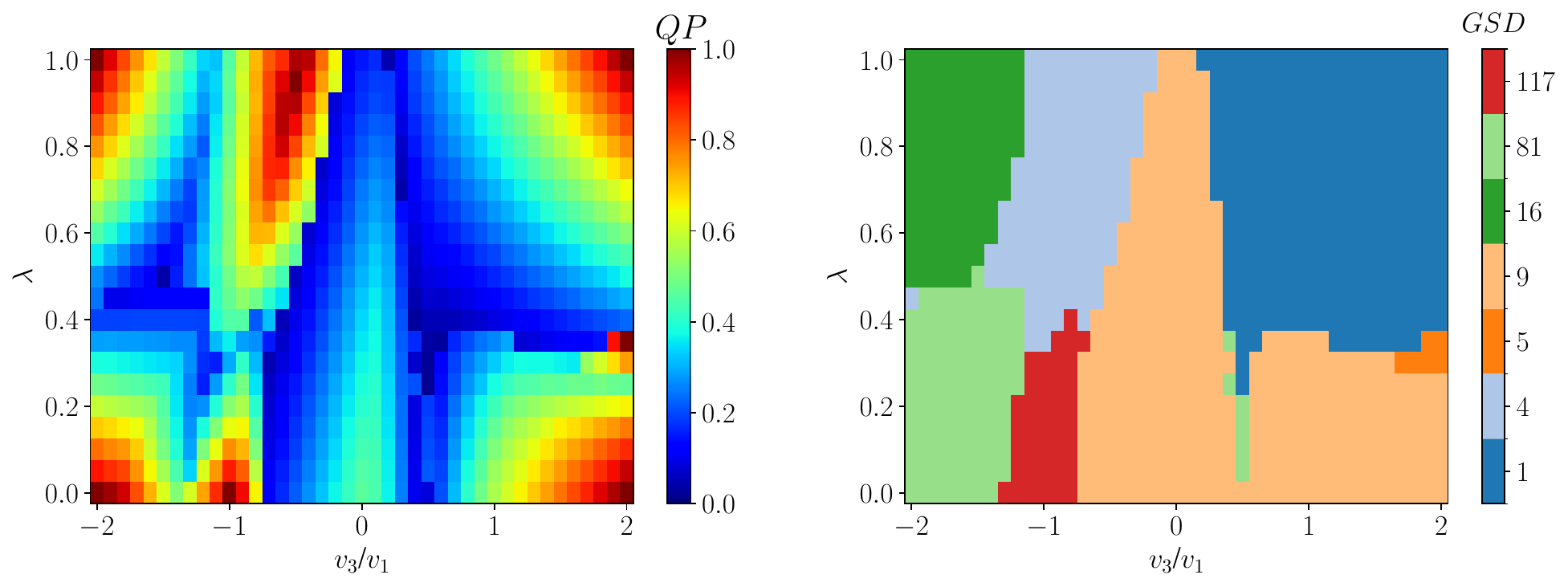}
    \caption{Phase diagram at fixed $v_0/v_1 =1.25$ in terms of the inter-spin coupling parameter $\lambda$ and $v_3/v_1$. Left: Once for given $(\lambda,v_m)$ configuration the gap for certain $\mathrm{GSD}$ is maximized, the phase is assumed to have said $\mathrm{GSD}$ and the quality parameter $\mathrm{QP}=\Delta_{\mathrm{GSD}}/\max_{\{\lambda,v_3\}}\Delta_{\mathrm{GSD}}$ is plotted accordingly in the corresponding region with associated amplitude. Right: Regions in the phase diagram marked by the respective detected $\mathrm{GSD}$. Featured are the PH(111) state ($\mathrm{GSD}=1$), the FTI ($\mathrm{GSD}=9$) and the $\mathrm{PS}$ I phase ($\mathrm{GSD}=4$). The other phases have not been investigated in the context of this work. The FTI at isotropic inter-spin coupling is adiabatically connected to the decoupled Laughlin limit and the paired state shows to require finite inter-spin correlation.}
    \label{fig:app9}
\end{figure}

\newpage

\section{Exact diagonalization in the sphere geometry}
\label{App3}

\noindent When maintaining the sign of the monopole charge in the center of the sphere, TR symmetry is achieved by particle-hole transforming the subsystem of spin-up electrons and flipping the spins to obtain the subsystem of spin-down electrons, and then flipping the signs of the inter-spin interactions. A $\nu_\mathrm{T}=2/3$ filled TR symmetric LLL for $N_\Phi=9$ then contains $N_\uparrow=4$ and $N_\downarrow=6$ electrons, respectively. In the $L_z=0$ angular momentum sector, the maximal angular momentum per species is $L_\mathrm{max}=N_\Phi N_\uparrow/2-N_\uparrow(N_\uparrow-1)/2=12$.

The projection and effect of the Haldane pseudopotentials, which are defined on the infinite $2$D plane, onto a system on the sphere, differs from the one onto a system on the torus. Hence, the amplitudes and phase boundaries will also differ.  Further, the notion of multifold ground state degeneracies is absent on the sphere. Phases that exhibit other identifying features, such as charge ordering, can be identified on the sphere and brought into comparison with the phase diagrams on the torus. Charge ordered phases are of non-zero ground state angular momentum and the phase can therefore be readily bounded by the investigation of the gap size at given pseudopotential configuration.

The charge density profiles shown in Figs.~\ref{fig:sphere3} and \ref{fig:sphere4} contain three plots, the first of which containing the density of spin up electrons, $\rho_\uparrow$, along the azimuthal direction in units of $r/l_B$ ($y$-axis) and given variation of a single pseudopotential ($x$-axis). In the center, the inverse particle-hole transformed electron density, $\rho_\downarrow=N_\mathrm{tot}-\Tilde{\rho}$, where $\Tilde{\rho}$ is the computed electron density for the spin down electrons, is shown. The reason is to better illustrate deviations in the behavior of the two species and bridge the gap to the description on the torus. The plot to the right shows the total electron density on the sphere, $\rho_\mathrm{tot}=\rho_\uparrow+\Tilde{\rho}_\downarrow$. Thus, simultaneous uneven distribution of spins on the sphere does not imply spin order, but indeed charge order.

\begin{figure}[htpb!]
    \centering
    \includegraphics[width=\linewidth]{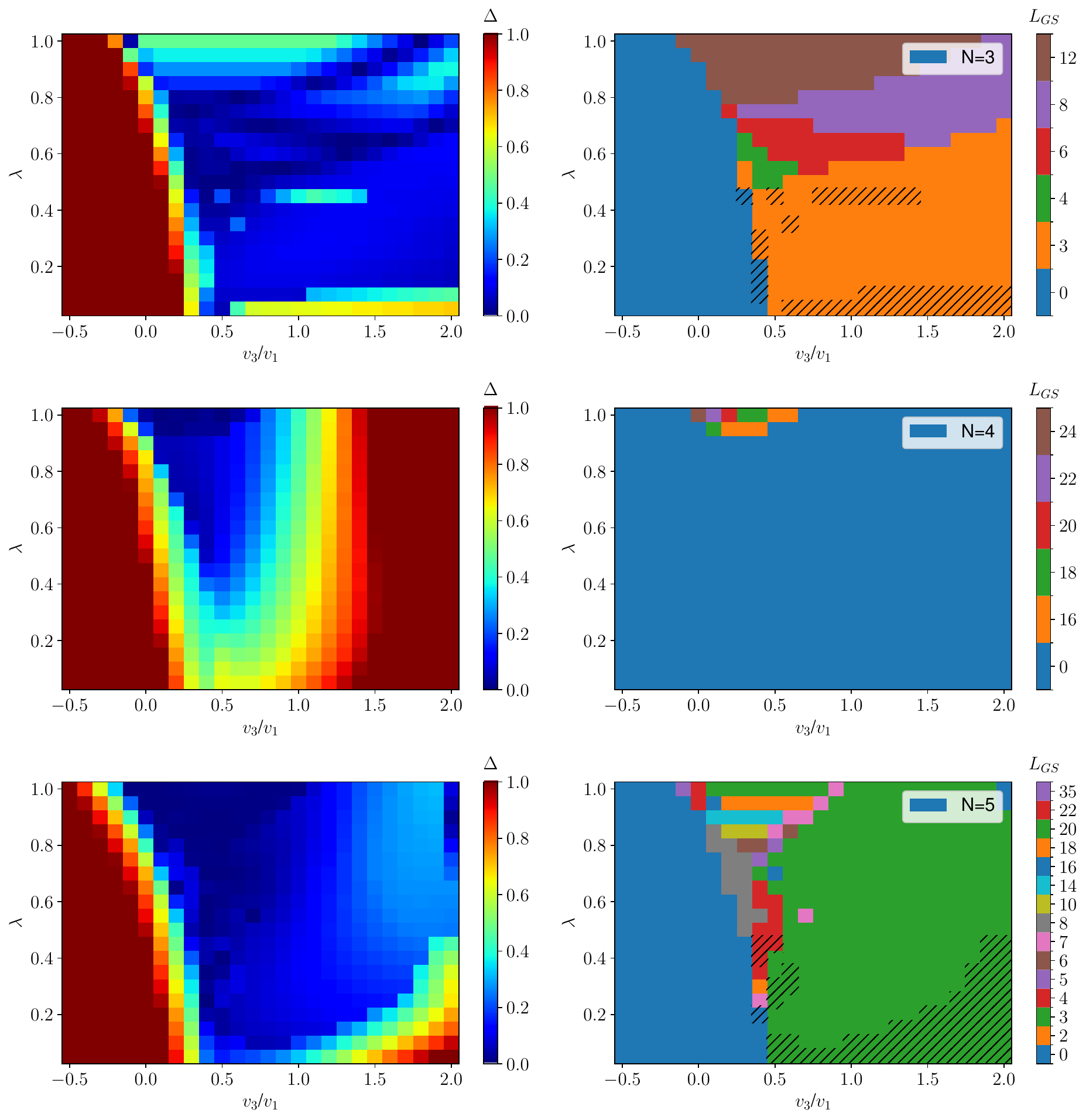}
    \caption{Left: Phase diagrams for the gap sizes at given particle number $N_\uparrow=N$ in terms of the repulsive interaction $-0.5<v_3/v_1<2$ and inter-spin coupling parameter $\lambda$. At $\lambda=0$ the spin species are completely decoupled. If the system is of angular momentum $L_\uparrow\otimes L_\downarrow=4$, the $L_z=0$ ground state sector features multiplets of $5$ states, which are completely degenerate in the $\lambda=0$ limit. For consistency, for regions of non-zero angular momentum the gap $E_5-E_4$ is considered. Right: Angular momentum of the ground state. If the system is of total angular momentum $4$, it is marked by hatching. It is evident that for $N=3,5$ the system at complete decoupling transitions into a phase of non-zero angular momentum, while remaining in the zero angular momentum phase for $N=4$. Inter-spin coupling does not alter the angular momentum, but hinders the detection by the loss of exact degeneracy and mixing with the excited spectrum.}
    \label{fig:sphere7}
\end{figure}

\begin{figure}[htpb!]
    \centering
    \includegraphics[width=0.75\linewidth]{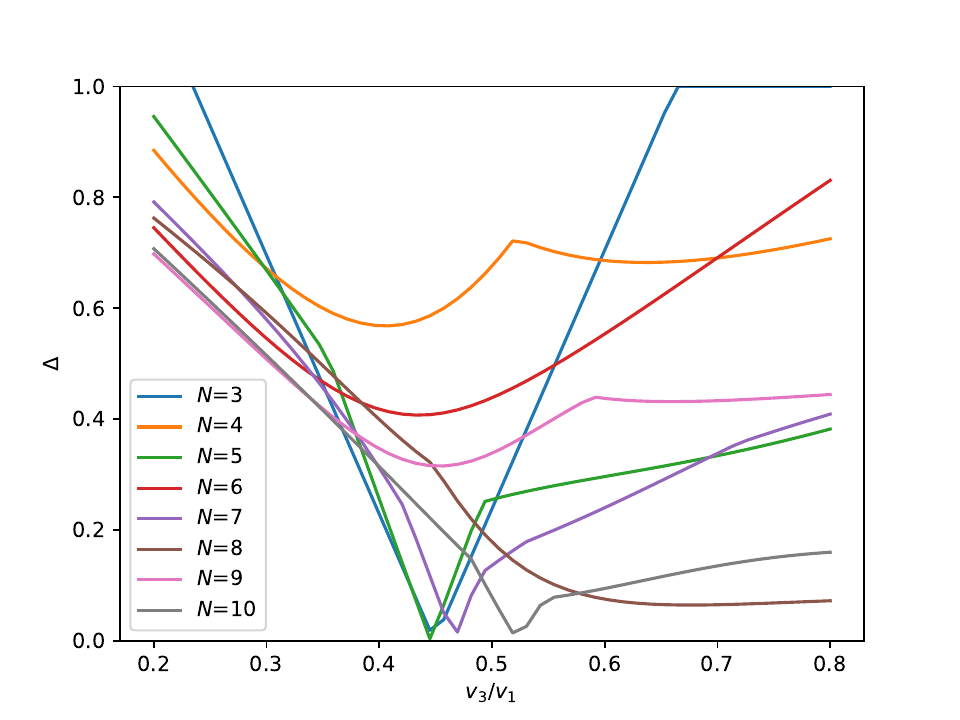}
    \caption{Phase diagram for the gap size $\Delta$ for a single spin species of varying particle number $N$ on the sphere. If a gap closing, i.e. $\Delta\sim0$, occurs, the system assumes $L=2$ by the argument in Fig.~\ref{fig:sphere7}. This transition occurs for $N=3,5,7,10$ and does not follow a clear distinguishable pattern in the set of probed particle numbers. Investigation of this phenomenon requires studies of higher particle numbers and was not further pursued as its nature as a finite-size effect unimportant in the thermodynamic limit.}
    \label{fig:sphere8}
\end{figure}

Our study further reveals that for a LLL of a single spin species there is a particle number dependence in whether or not the angular momentum of the system changes upon increment of repulsive interactions $v_3$. Fig.~\ref{fig:sphere7} shows a transition into a $L_\uparrow\otimes L_\downarrow=4$ angular momentum state for $N_\uparrow=3,5$ while remaining at zero for $N_\uparrow=4$. This phenomenon is further illustrated in Fig.~\ref{fig:sphere8}.

\begin{figure}[htpb!]
    \centering
    \includegraphics[width=\linewidth]{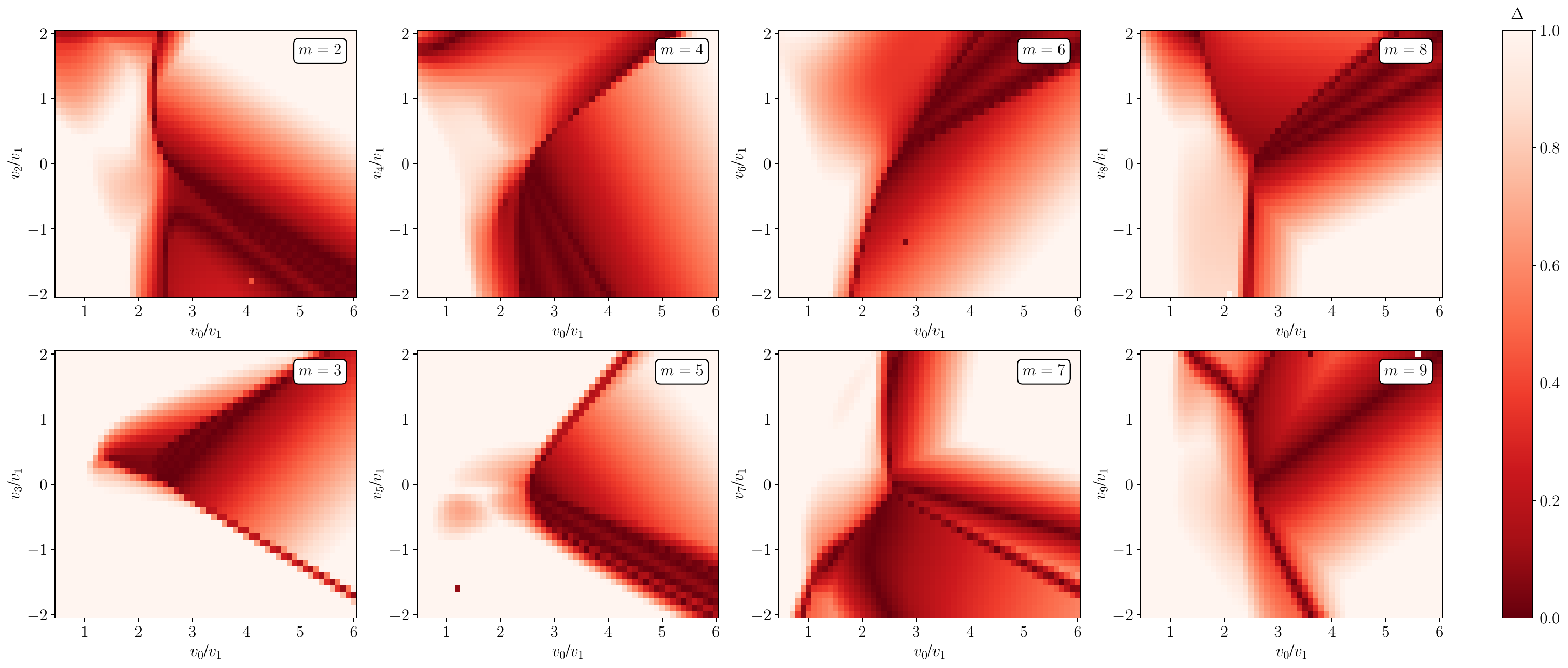}
    \caption{Phase diagrams for the gap sizes $\Delta=E_1-E_0$ between the ground state and the first excited state for given $(v_0,v_m)$ configuration. The system parameters are $N_\Phi =9$, $N_\uparrow=4$ and $N_\downarrow=6$. As only gap closings and, thus, $\Delta\rightarrow0$ are of importance, gap sizes $\Delta>1$ are identified with unity. For given $m$, there are two identifiable transitions. A transition with a single closing of the gap and a transition of multiple gap closings. Whether the direct or indirect transition appears for attractive or repulsive interactions depends on $m$, but does not follow any clear pattern. These transitions result in the increment of the ground state angular momentum upon increasing $v_0 $ and is discussed in more detail in Figs.~\ref{fig:sphere2} - \ref{fig:sphere6}. The phases to the right hand side of the phase diagrams therefore correspond to the broken spatial symmetric phases that have been identified on the torus. The other phases cannot be identified conclusively, as they all share the single ground state at zero ground state angular momentum and adiabatic connection to the decoupled limit. Especially the FTI is obscured by the absence of identifying features.}
    \label{fig:sphere1}
\end{figure}

\begin{figure}[htpb!]
    \centering
    \includegraphics[width=\linewidth]{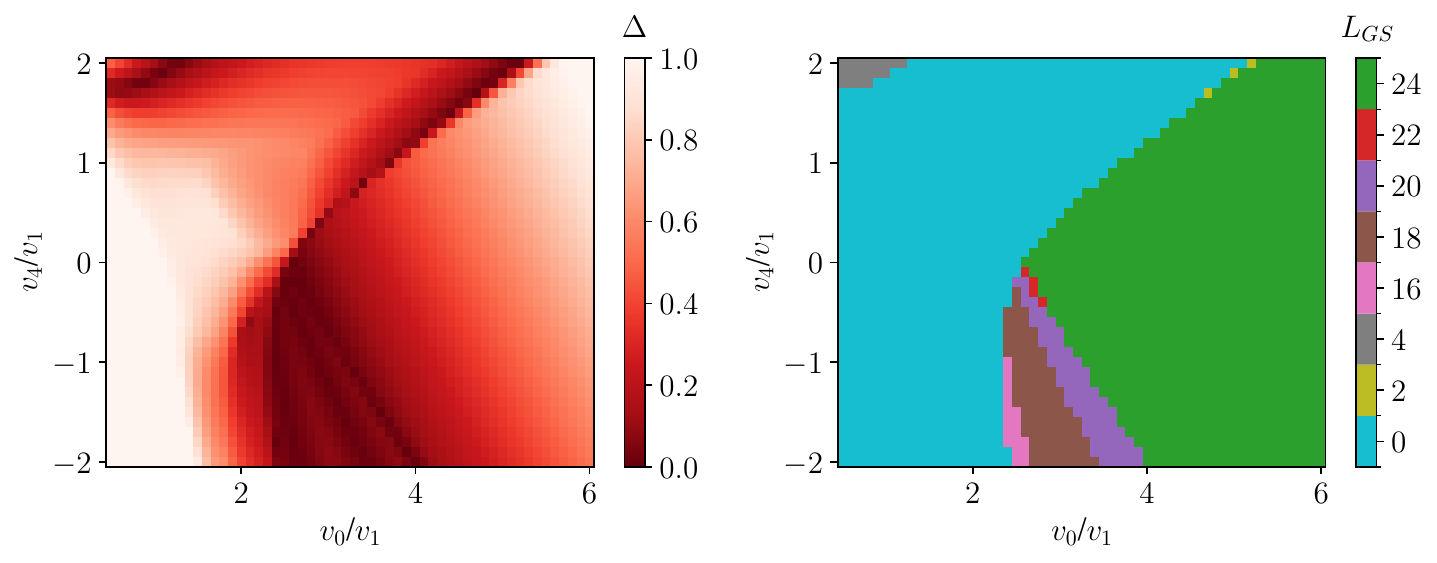}
    \caption{Left: The phase diagram for the gap $\Delta$ for $m=4$ for $N_\Phi =9$, $N_\uparrow=4$ and $N_\downarrow=6$. Right: The angular momentum of the ground state for given $(v_0,v_4)$ configuration, with $L_\mathrm{max}=24$ for the two species. The direct transition occurs at repulsive interaction, whereas the transition visiting intermediate states occurs at attractive interactions. Charge density profiles across the two transitions are shown in Figs.~\ref{fig:sphere3} and \ref{fig:sphere4}, and the respective spectra across these transitions in Figs.~\ref{fig:sphere5} and \ref{fig:sphere6}.}
    \label{fig:sphere2}
\end{figure}

\begin{figure}[htpb!]
    \centering
    \includegraphics[width=\linewidth]{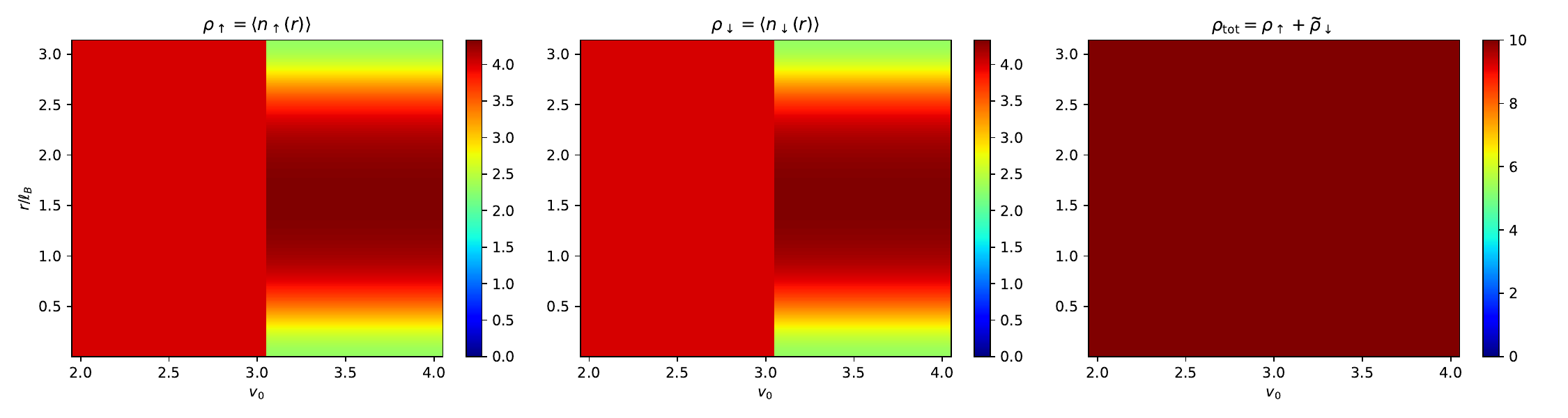}
    \caption{Electron density of the spin-up electrons (left), particle-hole transformed spin-down electrons (center) and the total density (right) across the direct transition in Fig.~\ref{fig:sphere2} at $v_4/v_1 = 0.5$ and $2<v_0/v_1<4$. Charge order is introduced after the gap is closed and the ground state angular momentum is maximized.}
    \label{fig:sphere3}
\end{figure}

\begin{figure}[htpb!]
    \centering
    \includegraphics[width=\linewidth]{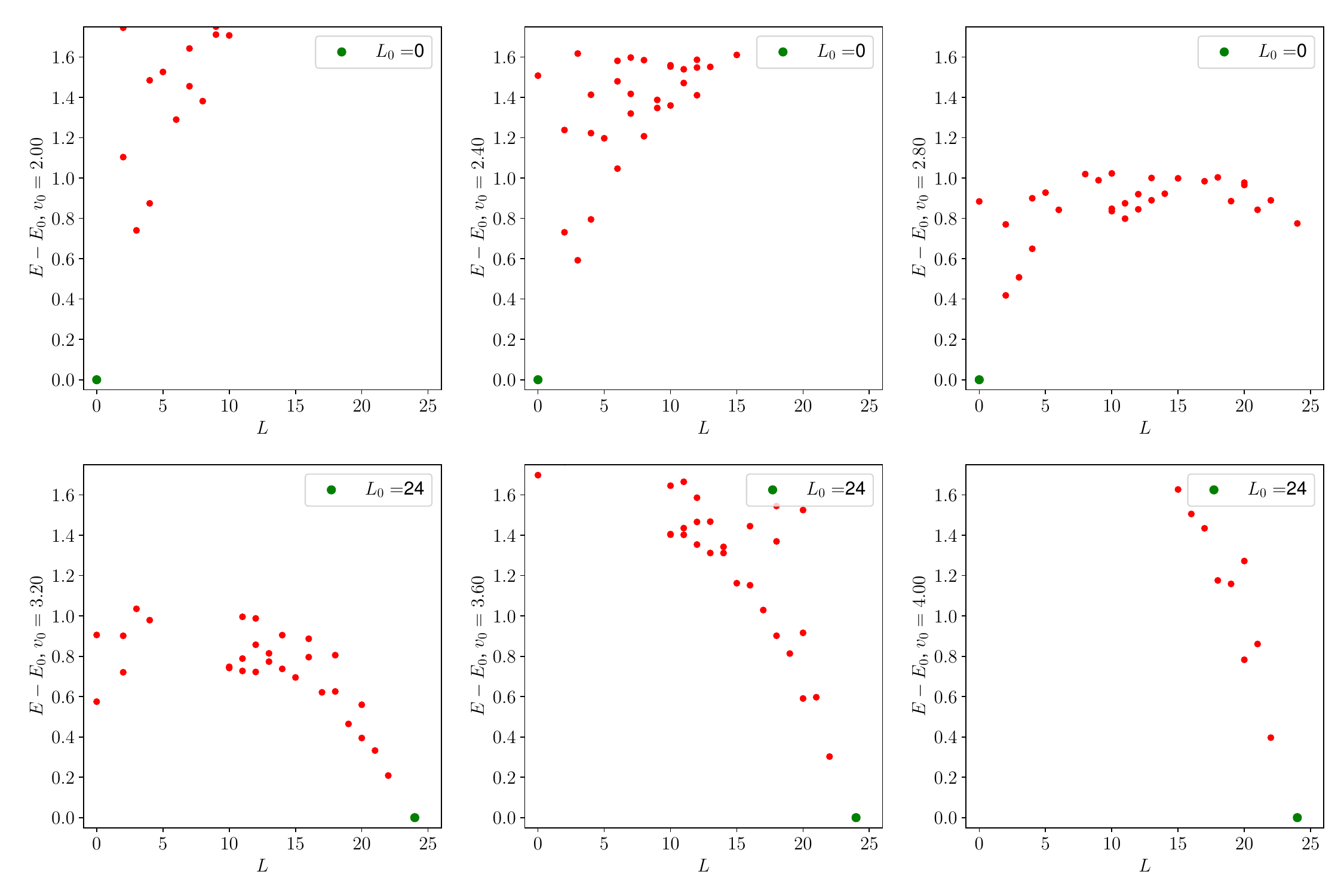}
    \caption{Spectra across the phase transition into the charge ordered state for the same system configuration as in Fig.~\ref{fig:sphere3}. The ground state is of zero angular momentum before the gap shrinks and an arch in the excited spectrum is visible. Finally, the state in the arch of maximal angular momentum minimizes the energy of the system and charge order is introduced. The spectra were computed in the $L_z=0$ sector.}
    \label{fig:sphere5}
\end{figure}

\begin{figure}[htpb!]
    \centering
    \includegraphics[width=\linewidth]{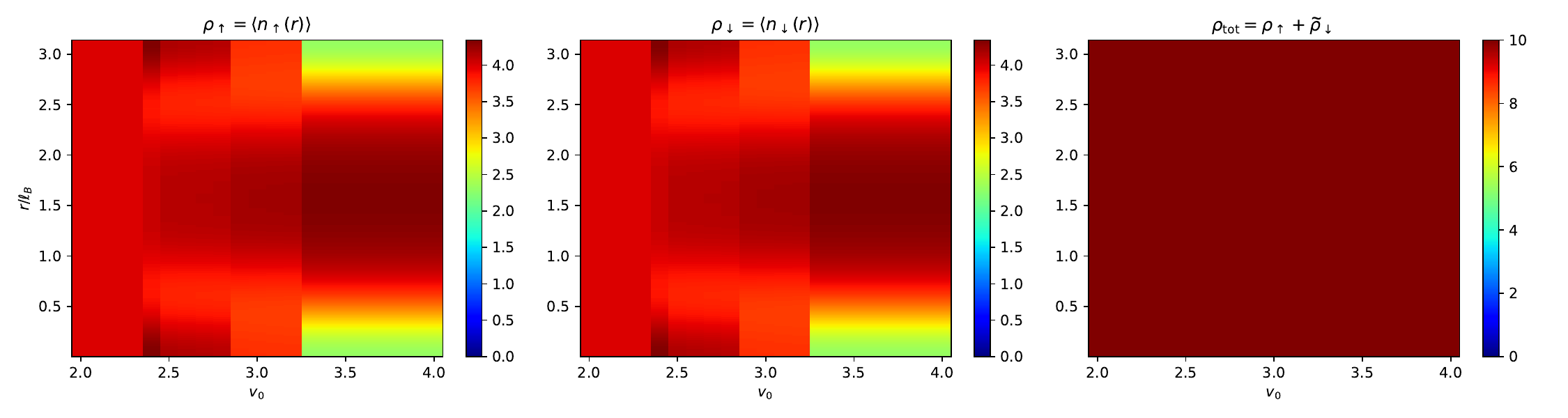}
    \caption{Electron density of the spin-up electrons (left), particle-hole transformed spin-down electrons (center) and the total density (right) across the direct transition in Fig.~\ref{fig:sphere2} at $v_4/v_1 = -1$ and $2<v_0/v_1<4$. Charge order is introduced after the first gap closure and the system is minimized by states of successively increasing angular momentum, before finally arriving at maximum angular momentum. While being present for any ground state angular momentum greater than zero, the charge order is most pronounced in the maximal angular momentum ground state.}
    \label{fig:sphere4}
\end{figure}

\begin{figure}[htpb!]
    \centering
    \includegraphics[width=\linewidth]{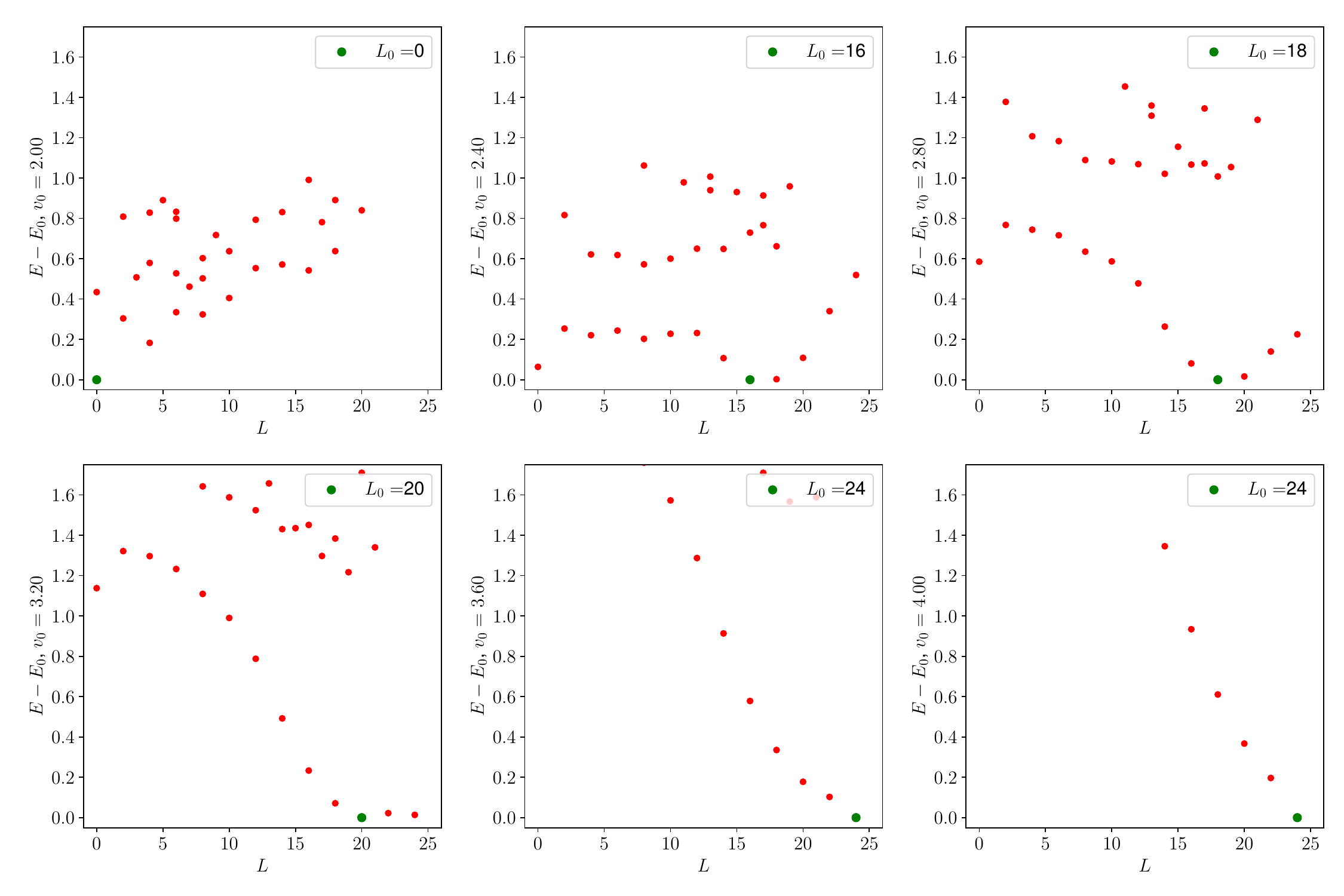}
    \caption{Spectra across the phase transition into the charge ordered state for the same system configuration as in Fig.~\ref{fig:sphere5}. The ground state is of zero angular momentum and relatively small gap before a distinct mode in the excited spectrum is visible. It is detached from higher excitations by a gap and shows a distinguishable shape. The minimum of this mode is the ground state and by the evolution of the system in $v_0$, the system is minimized by ground states of increasing angular momentum. Finally, the state of maximal angular momentum dominates and the gap increases. The spectra were computed in the $L_z=0$ sector.}
    \label{fig:sphere6}
\end{figure}

\end{appendix}

\end{document}